\documentclass[acmsmall]{acmart}
\usepackage{pifont}
\usepackage{hyperref}
\usepackage{hyperref}
\usepackage{algorithm} 
\usepackage[noend]{algpseudocode}
\algnewcommand{\Input}{\item[\textbf{Input:}]} 
\algnewcommand{\Output}{\item[\textbf{Output:}]} 
\usepackage{graphicx}   
\usepackage{subcaption} 
\usepackage{balance}
\usepackage{xcolor}
\hypersetup{
    colorlinks=true, 
    urlcolor=magenta,   
} 

\AtBeginDocument{%
  }

\copyrightyear{2026}
\acmYear{2026}

\setcopyright{cc}
\setcctype{by}
\acmJournal{PACMMOD}
\acmYear{2026} \acmVolume{4} \acmNumber{3 (SIGMOD)} \acmArticle{132}
\acmMonth{6} \acmDOI{10.1145/3802009}

\begin{document}



\title[AlignedServe: Orchestrating Prefix-aware Batching to Build a High-throughput \\ and Computing-efficient LLM Serving System]{AlignedServe: Orchestrating Prefix-aware Batching to Build a High-throughput and Computing-efficient LLM Serving System}

\author{Fengyao Bai}
\email{baify5@mail2.sysu.edu.cn}
\affiliation{%
  \institution{Sun Yat-Sen University}
  \city{Guangzhou}
  \country{China}
}
\author{Hongbin Zhang}
\email{zhanghb55@mail2.sysu.edu.cn}
\affiliation{%
  \institution{Sun Yat-Sen University}
  \city{Guangzhou}
  \country{China}
}

\author{Zhitao Chen}
\email{chenzht37@alumni.sysu.edu.cn}
\affiliation{%
  \institution{Sun Yat-Sen University}
  \city{Guangzhou}
  \country{China}
}
\author{Jiangsu Du}
\email{dujiangsu@mail.sysu.edu.cn}
\affiliation{%
  \institution{Sun Yat-Sen University}
  \city{Guangzhou}
  \country{China}
}

\author{Zhiguang Chen}
\authornote{Zhiguang Chen is the corresponding author.}
\email{chenzhg29@mail.sysu.edu.cn}
\affiliation{%
  \institution{Sun Yat-Sen University}
  \city{Guangzhou}
  \country{China}
}
\author{Yutong Lu}
\email{luyutong@mail.sysu.edu.cn}
\affiliation{%
  \institution{Sun Yat-Sen University}
  \city{Guangzhou}
  \country{China}
}

\renewcommand{\shortauthors}{Fengyao Bai et al.}


\begin{abstract}
High throughput inference serving is important for applications taking large language models (LLMs) as their kernels. However, traditional inference frameworks mostly suffer from the bubbles extensively existing in the inference pipeline. Research works have proposed to group multiple requests into batches and schedule these batches efficiently thus reduce the request-level and batch-level bubbles, but rarely pay attention to the bubbles within each decode iteration. Actually, tokens generated in the same iteration may have different costs depending on their relied KVCache, where a token relying on a very long KVCache is likely to be the bottleneck within the iteration, and consequently the iteration-level bubbles occur since other tokens must wait for a long time to enter into the next iteration.

In this work, we propose a novel prefix-aware batching policy to group requests whose relied KVCache are of the similar length into a batch, guaranteeing that bubbles within each iteration are eliminated. To efficiently support the prefix-aware batching, we design a new inference framework called AlignedServe, which leverages the large CPU memory to accommodate a sufficient amount of in-flight requests prepared for being batched. Batches generated in CPU memory are further scheduled by a well-designed batch-level scheduling policy, which guarantees that the batch-level bubbles are significantly reduced. To reduce the latency involved in transmitting KVCache from CPU memory to GPU HBM, we propose to leverage one GPU to prefetch KVCache for another. To the best of our knowledge, this is the first work employing the GPU-Prefetch-For-GPU architecture. We evaluate AlignedServe via extensive experiments driven by both synthetic and application workloads. The experimental results demonstrate that AlignedServe improves the decoding throughput by a maximum of 1.98$\times$ and reduces the latency by up to $7.4 \times$ compared to the state-of-the-art systems.
\end{abstract}
\ccsdesc[500]{Information systems~Data management systems}

\keywords{Large Language Models, Inference, Prefix-aware batching, Batch-level scheduling, High throughput inference}

\received{October 2025}
\received[revised]{January 2026}
\received[accepted]{February 2026}


\maketitle

\section{Introduction}
Large Language Models (LLMs) are extensively considered as the most attractive AI technology for the time being and have revolutionized a wide variety of applications, such as chatbots~\cite{google_gemini_2024,openai_chatgpt}, programming assistants~\cite{code_llama_2023,github_copilot} and text summarization~\cite{kimi2024,lewis-etal-2020-bart}. Driven by the increasing demands, larger and larger LLMs have been published as these powerful models are able to serve for more general-purpose scenarios. 
However, larger models introduce great challenges into both the training and inference due to their significant computing overhead. 
In this work, we focus on the inference which is more cost-sensitive since only the cost-effective models would be commercially successful. 

The LLM inference can be divided into two phases, i.e., the prefill and decode, where the prefill processes the input prompt as well as produces the first output token, and the decode generates subsequent tokens one by one until that the inference finishes. Among the two phases, the prefill is characterized to be computing-intensive thus is friendly to GPUs, whereas the decode relies on large volumes of KVCache~\cite{kitaev2020efficient} to compute the attention thus is generally memory-bound. Specifically, to generate a token, we must compute the attention from the Key-Value pairs of the entire input prompt as well as the tokens that have already been generated in the past decode iterations. Hereafter, we name the relied Key-Value pairs as the \textit{Prefix} of the token to be generated. 

As the prefill and decode present different workload characteristics, a large number of works have been proposed to facilitate the graceful cooperation between them. Generally, there are two concerned issues that should be considered seriously. 
The first is how to handle the discrepancy between prefill and decode. An intuitive solution about this issue is splitting the two phases onto separate GPUs, or even separate machines. Related works such as Splitwise~\cite{patel2024splitwise}, DistServe~\cite{zhong2024distserve}, Mooncake~\cite{305212}, D{\'e}j{\`a}Vu ~\cite{stratidejavu} etc belong to this category, and have succeeded in improving the overall performance of inference significantly. However, splitting prefill and decode onto separate GPUs presents a dilemma, as it is difficult to balance the workloads among GPUs assigned to different tasks. As a result, some GPUs are inevitably underutilized.


The second concerned issue is how to improve the computing intensity of the decode iterations. As the decode iterations are mostly memory-bound, GPUs responsible for decode are unlikely to be saturated under limited concurrent token generation tasks. Accordingly, these state-of-the-art works, e.g. Orca~\cite{orca}, TetriInfer~\cite{hu2024inference}, Sarathi-Serve~\cite{sarathi-serve}, LoongServe~\cite{wu2024loongserve}, all propose to group tens or hundreds of inference requests into a batch, and schedule these batches among GPUs to achieve higher computing efficiency thus higher throughput. However, we argue that existing batching policies mostly target to group as large batches as possible to enhance computing intensity, and existing scheduling policies mostly focus on efficiently dispatching these batches among GPUs to eliminate bubbles~\cite{feng2023mobius,narayanan2019pipedream}. They optimize the computing efficiency at the coarse-grained request-level (i.e., grouping multiple requests into a batch) and batch-level (i.e., dispatching multiple batches among GPUs), respectively, but have not shed any light on the optimization at the fine-grained iteration-level. Actually, we find that the bubbles within each iteration extensively exist, as discussed below. 

In a given iteration, each request within the batch generates a new token. To the best of our knowledge, there are no research works considering about the different costs of generating tokens of different requests in an iteration.
Generally, the cost of generating a token consists of two components, computing the MLP (Multilayer Perceptron) and attention, where the cost introduced by MLP is identical for every token, while the cost introduced by attention is partially determined by the prefix of the token has been generated, indicating that tokens generated in the same iteration have different costs since their prefix may be of different lengths. As a result, the token with a very long prefix is likely to be the bottleneck within the iteration, and accordingly bubbles at the iteration-level occur as tokens with short prefix are produced quickly and must wait for the generation of tokens with long prefix. The above conclusion will be demonstrated by experiments in Section~\ref{section2}. 

Unfortunately, existing batching policies have rarely taken the bubbles within iterations into account. 
For example, the state-of-the-art Orca ~\cite{orca} proposed the iteration-level scheduling (known as continuous batching), which permits a request to leave from the batch upon its completion, and accordingly accepts a new request immediately to fill the free slot. By doing so, the bubbles at the request-level are eliminated since the batch always maintains a sufficient number of requests. Similarly, Sarathi-Serve~\cite{sarathi-serve} and DeepSpeed-FastGen ~\cite{holmes2024deepspeed} also pursued to generate as large batches as possible by composing tokens from both prefill and decode. None of the state-of-the-art works mentioned above have paid special attention to the discrepancy between tokens generated in each given iteration.

Motivated by the challenges in terms of the multi-level bubbles as discussed above, in this work, we propose a novel LLM inference framework as well as the corresponding prefix-aware batching and batch-level scheduling policy to eliminate bubbles at different levels thus achieve high throughput. Contributions of our work can be summarized as follows.

\begin{itemize}
    \item We give a deep insight into the decode within each iteration via both theoretical analysis and trace-driven experiments, demonstrating that tokens generated in the same iteration may have different costs depending on the lengths of their prefixes. The disparity between tokens introduces iteration-level bubbles, which will degrade the overall performance significantly. To the best of our knowledge, this is the first work considers iteration-level bubbles. 

    \item We propose a prefix-aware batching policy which pursues to group requests with the similar length of prefix into a batch, guaranteeing that all tokens generated in an iteration have the same cost, and consequently bubbles within iterations are eliminated. Note that, the prefix of a token includes both the input prompt and tokens that have already been generated. It is actually the KVCache accumulated by the request.

    \item We design a novel inference framework as well as the corresponding batch-level scheduling policy to efficiently support the prefix-aware batching. Specifically, the framework decouples the prefill and decode phases by assigning them to separate GPUs, and exploits the large CPU memory to keep the KVCache of these in-flight inference requests. The incoming inference requests are first processed by prefill GPUs, where the obtained KVCache would not be transmitted to decode GPUs immediately, but are offloaded to the large CPU memory instead. Requests accumulated in CPU memory are grouped by the prefix-aware batching policy into batches, which are scheduled to decode GPUs by our batch-level scheduling policy ultimately. Note that, the KVCache are not directly transmitted from CPU memory to decode GPUs via PCIe interface, but are prefetched to prefill GPUs beforehand, and then transmitted to decode GPUs via the high-performance NVLink express. As the latency of transmitting KVCache via NVLink is much lower than that of PCIe, the batch-level bubbles will be significantly reduced further. To the best of our knowledge, this is the first work that leverages prefill GPUs to prefetch KVCache for decode GPUs. As the requests belonging to the same batch are aligned in terms of prefix length, we name our framework AlignedServe.         

    \item We evaluate AlignedServe via extensive experiments driven by both synthetic and application workloads. The experimental results demonstrate that AlignedServe improves the decoding throughput by a maximum of 1.98$\times$ and reduces the latency by up to $7.4 \times$ compared to the state-of-the-art systems.
\end{itemize}

\section{Background and Motivation}   \label{section2} 

In this section, we describe the decoder-only Transformer~\cite{vaswani2017attention} LLMs along with the auto-regressive inference. Based on a deep insight into the inference process driven by both theoretical analysis and experiments, we demonstrate that bubbles within each iteration are non-trivial.  

\subsection{LLM Inference}  

Popular LLMs such as GPT~\cite{10.5555/3495724.3495883}, Llama~\cite{DBLP:journals/corr/abs-2302-13971}, OPT~\cite{DBLP:journals/corr/abs-2205-01068} etc. mostly adopt the decoder-only Transformer architecture and use the auto-regressive method to generate output tokens. Specifically, the LLM inference can be divided into two phases: \textit{Prefill} and \textit{Decode}.

\textbf{Prefill phase:} For a given inference request, the prefill phase accepts the whole input prompt and computes all the tokens in parallel to generate the first output token. The intrinsic parallelism indicates that the prefill is computing-efficient and introduces into relatively low overhead. As demonstrated by Agrawal et al.~\cite{agrawal2023sarathi}, the cost associated with the prefill phase can be as low as $1/200$ of the cost of decoding at small batch sizes.

\textbf{Decode phase:} After the prefill phase, the decode phase generates subsequent tokens one by one until the inference completes. According to the auto-regressive model, the generation of a given token depends on all the previous tokens that have already been generated in both the decode phase and prefill phase. Once a token is generated, it is appended to the end of the token sequence as well, contributing to the generation of the following tokens. As each token will repeatedly join the subsequent generation tasks, an intuitive optimization is keeping the key and value vectors used to characterize the token in memory. This data structure is known to be KVCache~\cite{vllm}, which has been extensively used in LLM inference to reduce computational overhead. 

Compared with the prefill phase, the decode phase is less efficient for the following two reasons. First, the decode phase presents poor parallelism since tokens are generated one by one. Although grouping hundreds of requests into a batch helps to improve the computing intensity, the required memory usually overwhelms the HBM deployed in GPUs. Second, the decode phase is memory-bound thus is unlikely to saturate the computing power of GPUs. Considering the above challenges, researchers argue that the overhead of LLM inference is mostly introduced by the decode phase, which is the major concern of this work.

\subsection{A Deep Insight into the Decode}   \label{subsection2.2}
A typical Transformer model consists of $l$ identical layers, each of which is divided into two parts: a Multi-Head Attention (MHA) block and the Multi-Layer Perceptron (MLP) block. To describe these two types of blocks, We define the following notations: the parameter $b$ denotes the batch size, the parameter $s$ denotes the length of an inference sequence, the parameter $h$ denotes the hidden dimension. Based on the defined parameters, the computational overheads of MHA and MLP can be estimated as follows, respectively.

\textbf{Computational Overhead of MHA.} For a new token $x_{n+1}$ to be generated, we assume that the prefix of $x_{n+1}$ is $X = {x_1, ... ,x_n }$, where $x_i$ is either contained in the prompt or generated in the decode phase. As KVCache has been extensively adopted by LLM inference, we actually do not save $X$, but keep $K = {k_1, ... ,k_n }$ and $V = {v_1, ... ,v_n }$ in memory, where both $k_i$ and $v_i$ are computed by the matrix-vector multiplication in the form of $[h,h]*h$ according to Formula~\ref{equation1}. 

\begin{equation}
q_i = W_{q}x_i,\ k_i = W_{k}x_i,\ v_i = W_{v}x_i \label{equation1}
\end{equation}

To generate the token $x_{n+1}$, we must compute the attention between the token $x_{n}$ and the prefix of $x_{n+1}$, i.e., $X = {x_1, ... ,x_n }$ according to Formula~\ref{equation2}. The kernel of Formula~\ref{equation2} is two vector-matrix multiplications in the form of $h\times[h,s]$ and $s\times[s,h]$, respectively. The total number of computing operations involved in Formula~\ref{equation2} can be approximately estimated as $2sh$, where the computational overhead introduced by $softmax$ is mostly considered to be negligible. The involved data to be accessed is the KVCache, which can be estimated as $2sh$, where the number $2$ indicates two types of vector (i.e., $k$ and $v$), $s$ is the number of vectors in each type, and $h$ is the size of a vector.

\begin{equation}
Attention(q,K,V) = softmax\left(\frac{qK^{T}}{\sqrt{d_{k}}}\right)V\label{equation2}
\end{equation}

\textbf{Computational Overhead of MLP.} After the attention has been computed, the obtained vector (with size $h$) is fed to the MLP which consists of two fully-connected layers. The first layer expands the feature dimension to enhance the representational capacity, and the second layer restores the feature dimension and serves as the output layer. Formula~\ref{equation3} presents the MLP in detail, where $W_1$ and $W_2$ are in the shape of $[h,4h]$ and $[4h,h]$, respectively. 

\begin{equation}
x = f_{GeLU}(x_{out}W_{1})W_{2}+x_{out}\label{equation3}
\end{equation}

According to Formula~\ref{equation3}, the two layers conduct vector-matrix multiplication in the form of $h\times[h,4h]$ and $4h\times[4h,h]$, respectively. Therefore, the total number of computing operations involved in these layers can be approximately estimated as $8h^2$, where the computational overhead introduced by $f_{GeLU}$ is mostly considered to be negligible. The involved data to be accessed is the two weight matrices $W_1$ and $W_2$, which can be be estimated as $8h^2$.

Table \ref{table1} summarizes the computational and memory overheads involved in the generation of a single token. A straightforward observation from the table indicates that the overhead of MLP is much higher than that of MHA in terms of both computing and memory. However, we argue that the performance of MLP will be significantly improved upon large batches since the weight matrices $W_1$ and $W_2$ are shared by all tokens within a batch, the overhead of loading $W_1$ and $W_2$ from HBM will be amortized among tens or even hundreds of tokens. On the contrary, the MHA slightly benefits from large batches since the overhead is mostly introduced by KVCache, which is linearly increased with respect to the batch size. Furthermore, as more and more tokens have been generated within a sequence, the overhead of MLP remains constant, while the overhead of MHA increases gradually with respect to the parameter $s$. In conclusion, the MHA contributes a large fraction of the overall overhead, especially when the token to be generated relies on a long prefix. Based on the above analysis, we consider two tokens: one that relies on a very long prefix and the other on a relatively short prefix; their generation overheads differ substantially. If they are generated in the same iteration within a batch, the token introducing higher overhead becomes the bottleneck and significantly degrades the overall performance, as demonstrated by experiments in the next subsection.

\begin{table}[t]
  \caption{The overheads of generating a single token}
  \label{table1} 
  \begin{tabular}{ccc}
    \toprule
    Component&Computational Overhead&Memory Overhead\\
    \midrule
    MHA & $2sh$& $2sh$\\
    MLP & $8h^2$& $8h^2$\\
  \bottomrule
\end{tabular}
\end{table}
\subsection{The Bottleneck of Tokens with Long Prefix} \label{section2.3}

\begin{figure*}[t] 
    \centering 
    \begin{subfigure}[t]{0.248\textwidth} 
        \centering
        \includegraphics[width=\textwidth]{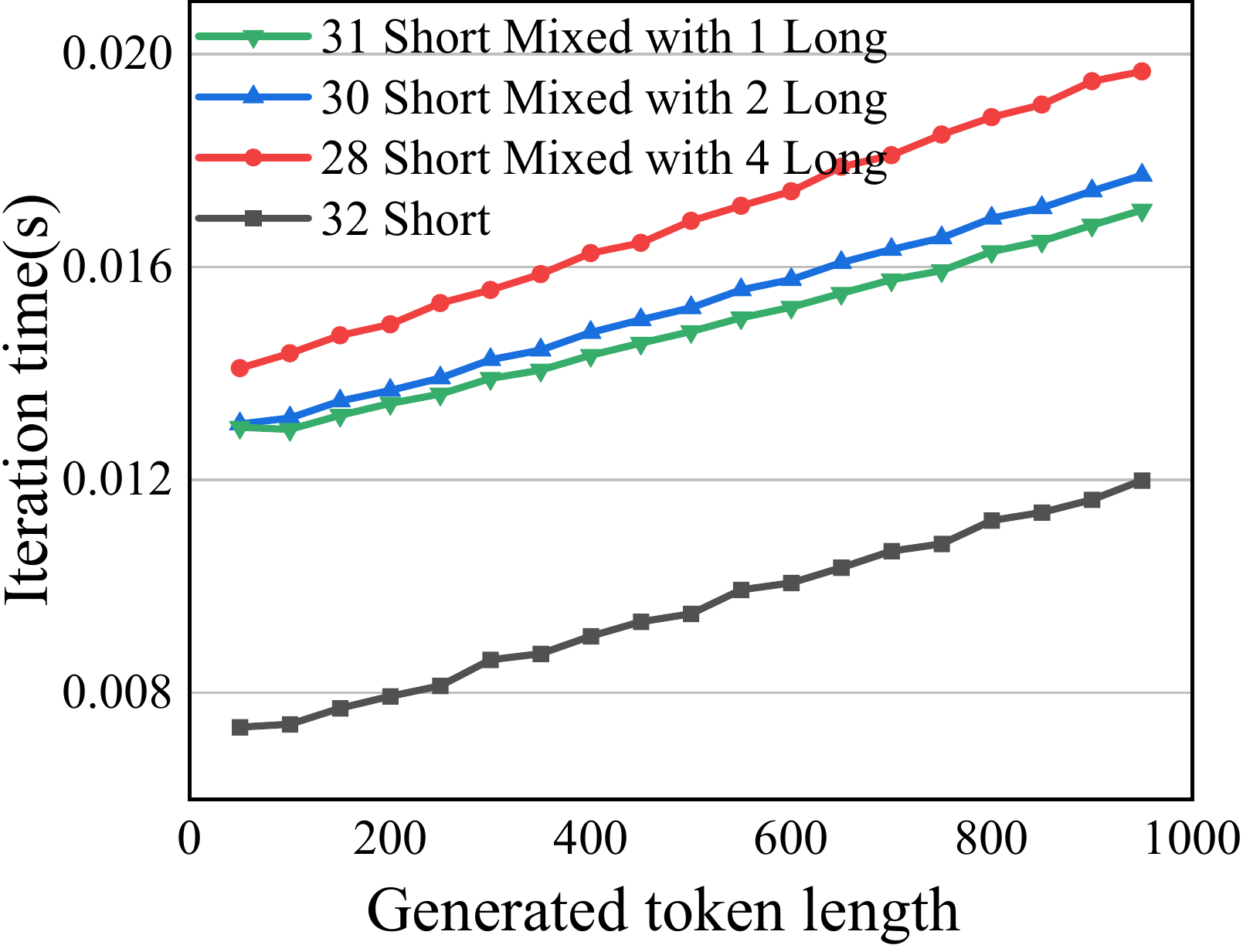}
        \caption{Batch size 32.}
    \end{subfigure}%
    \begin{subfigure}[t]{0.248\textwidth}
        \centering
        \includegraphics[width=\textwidth]{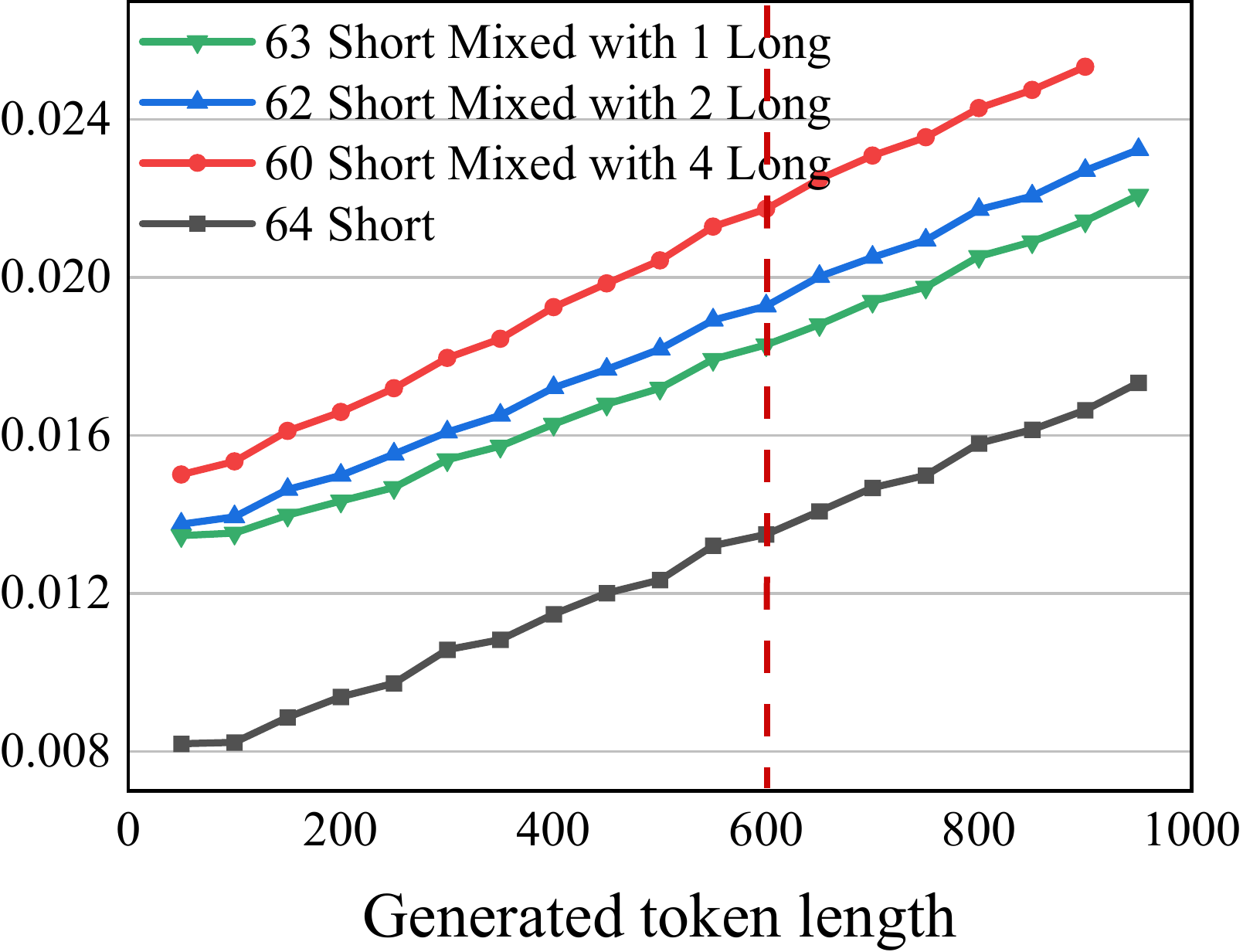}
        \caption{Batch size 64.}
    \end{subfigure}
    \begin{subfigure}[t]{0.248\textwidth}
        \centering
        \includegraphics[width=\textwidth]{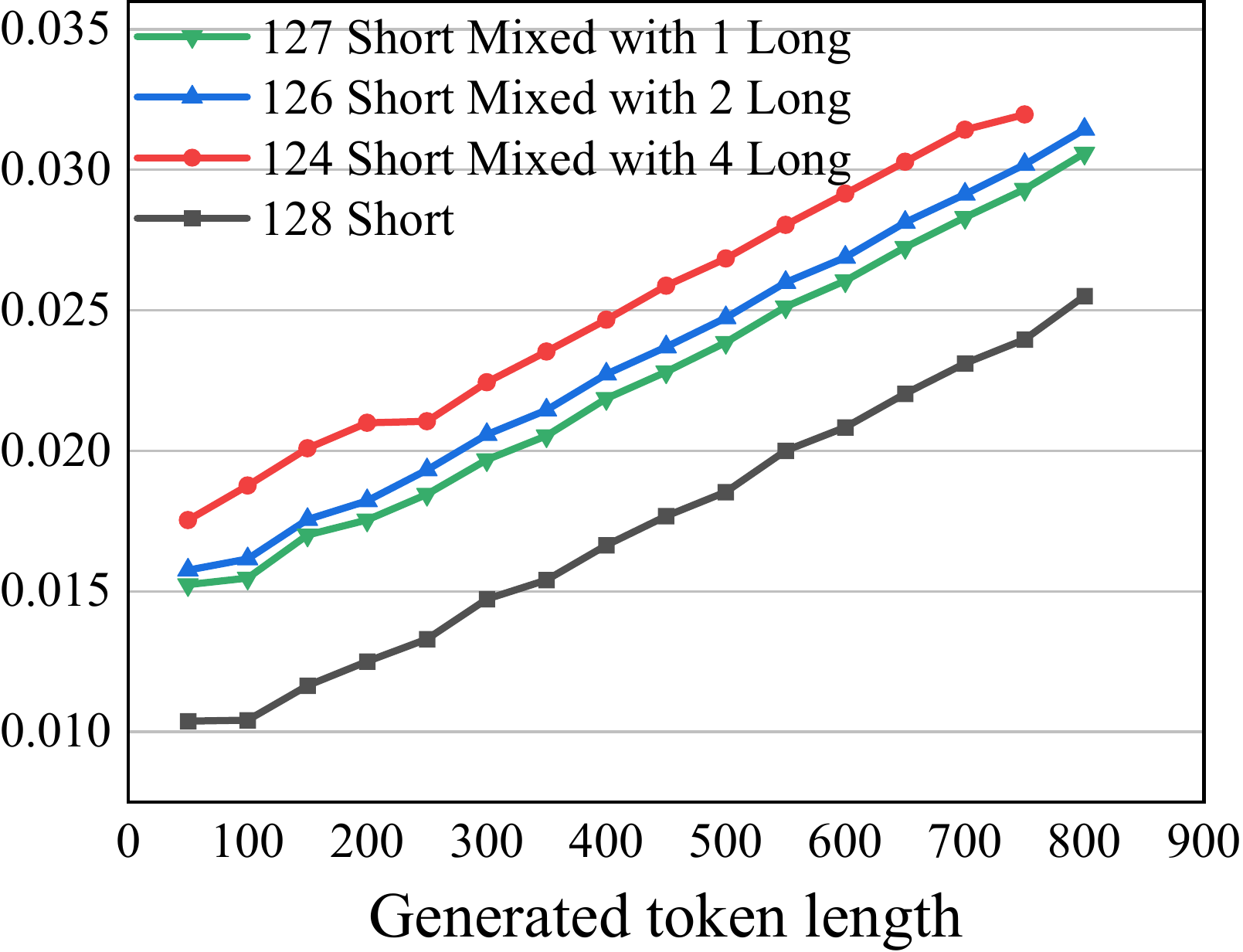} 
        \caption{Batch size 128.}
    \end{subfigure}%
    \begin{subfigure}[t]{0.248\textwidth}
        \centering
        \includegraphics[width=\textwidth]{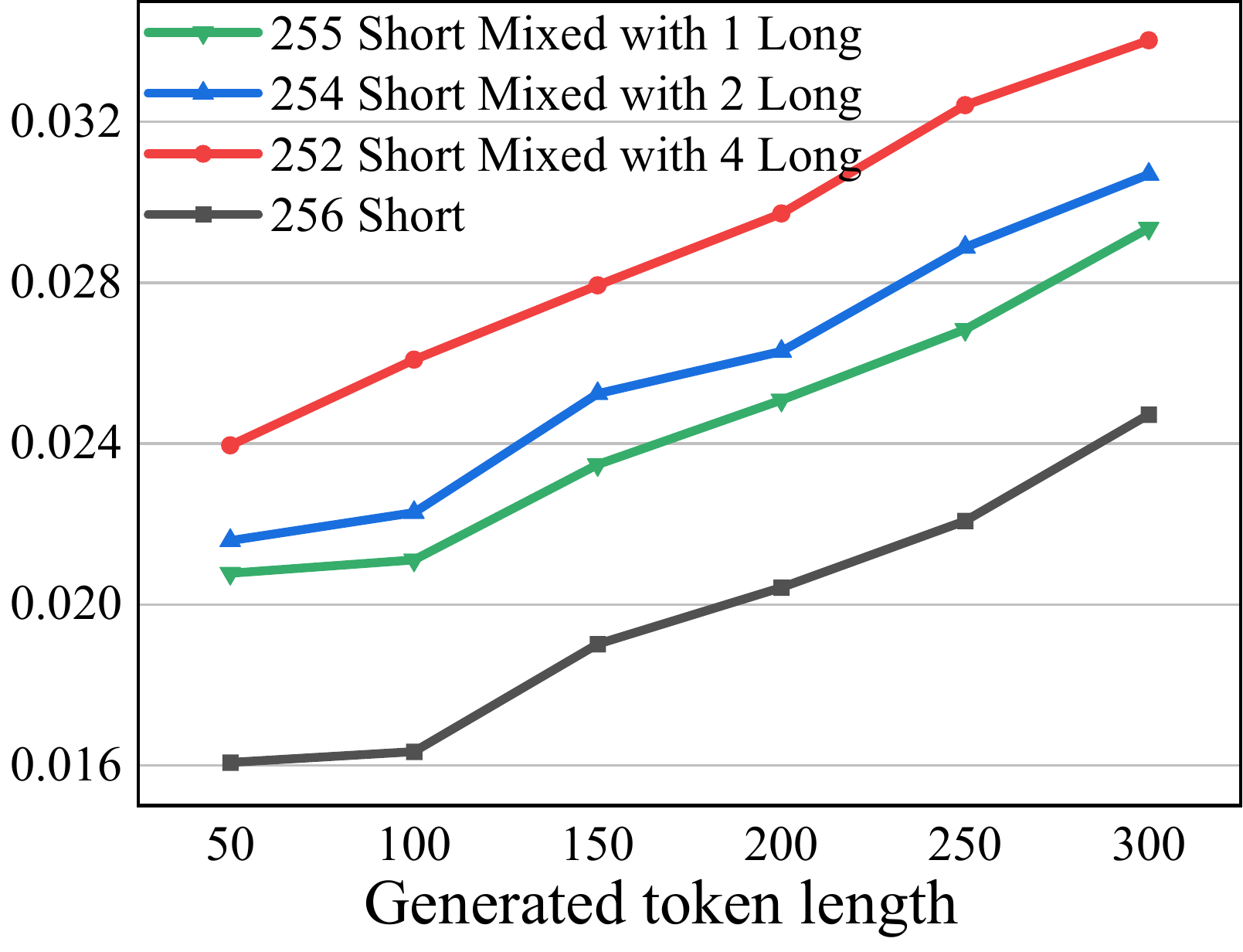}
        \caption{Batch size 256.}
    \end{subfigure}
    \caption{The negative impact introduced by the tokens with long prefix in each iteration.}
    \label{motivation1}
\end{figure*}

To demonstrate that tokens with long prefix will become the bottleneck in each iteration, we design experiments which group different lengths of prompts into a batch, and evaluate the latency of each iteration. Specifically, in each experiment, we compare the latency of each iteration under four types of batches. Taking the batch size 64 as an example, the baseline is a batch containing 64 short prompts, where each prompt contains only 32 tokens. The other three comparison candidates are 63 short prompts mixed with 1 long prompt, 62 short prompts mixed with 2 long prompts, 60 short prompts mixed with 4 long prompts, respectively, where the long prompt contains as many as 4096 tokens. We measure the latency of each iteration and verify the nontrivial negative impact introduced by these long prompts, even though the long prompts contained in each batch are no more than 4. All the experiments run Llama-7b model~\cite{Schmid2023} on vLLM~\cite{vllm2024} deployed on H100 GPUs. Experimental results are presented in Figure~\ref{motivation1}.

From Figure \ref{motivation1} we make two observations. The first is that as more and more tokens have been generated by decode, the latency of each iteration increases gradually since the relied prefix become longer. This observation indicates that the long prefix does impose significant impact on the latency of decode, which conforms to the conclusion derived from the theoretical analysis in Subsection~\ref{subsection2.2}. The second observation is that even a batch contains few long prompts, the overall performance will be degraded remarkably. Taking the batch size 64 as an example, when the generated token length is 600 (marked by the vertical red dotted line), the latencies of four comparison candidates (i.e., the baseline, 63 short prompts mixed with 1 long prompt, 62 short prompts mixed with 2 long prompts, 60 short prompts mixed with 4 long prompts) are $13.49ms$, $18.29ms$, $19.27ms$, $21.73ms$, respectively, where only 4 long prompts will increase the latency of an iteration by about $61\%$ (i.e., $(21.73-13.49)/13.49$).  

\begin{figure}[t]
    \begin{subfigure}[t]{0.3\linewidth}
        \centering
        \includegraphics[width=\textwidth]{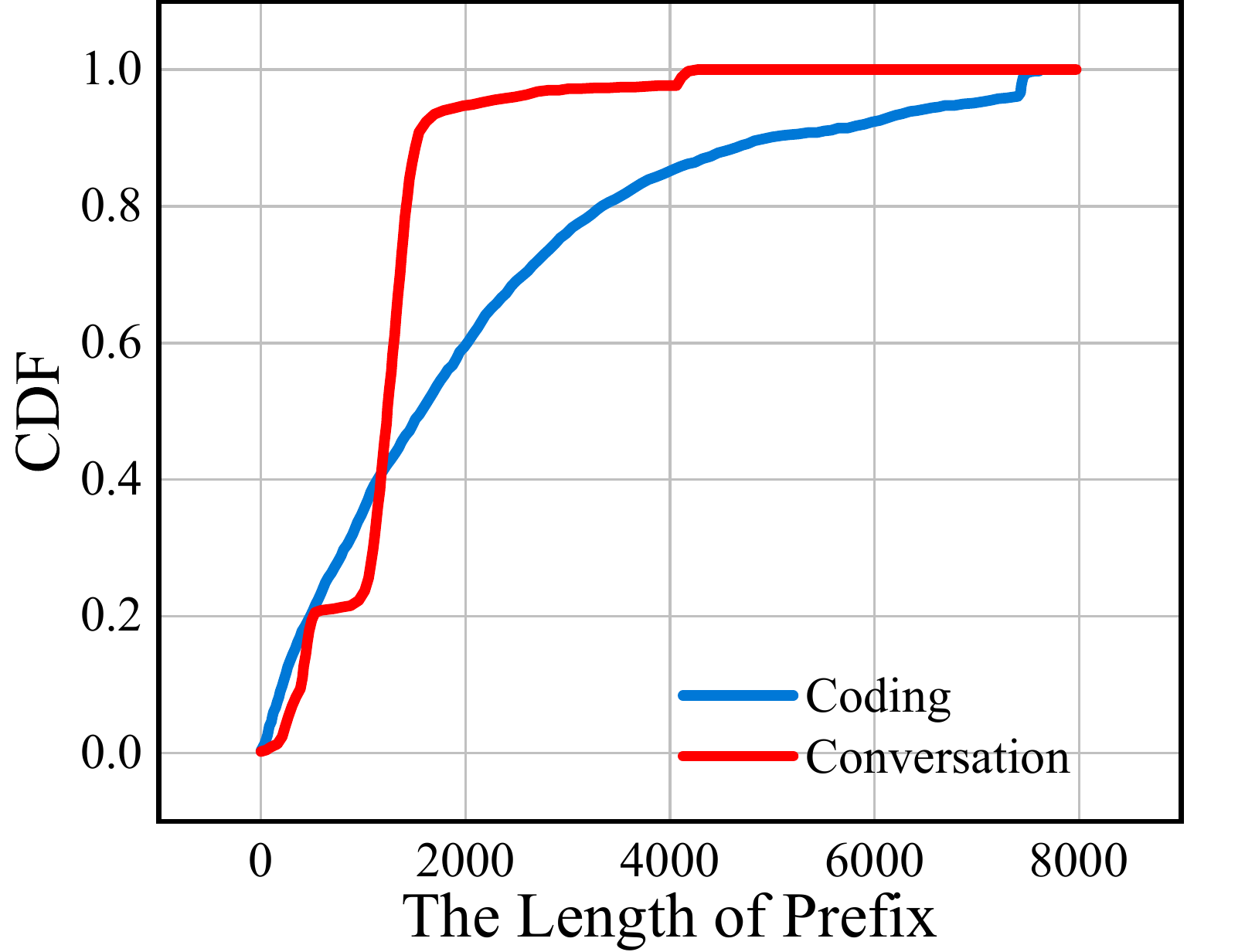} 
        \caption{AzurePublicDataset.}
    \end{subfigure}%
    \begin{subfigure}[t]{0.3\linewidth}
        \centering
        \includegraphics[width=\textwidth]{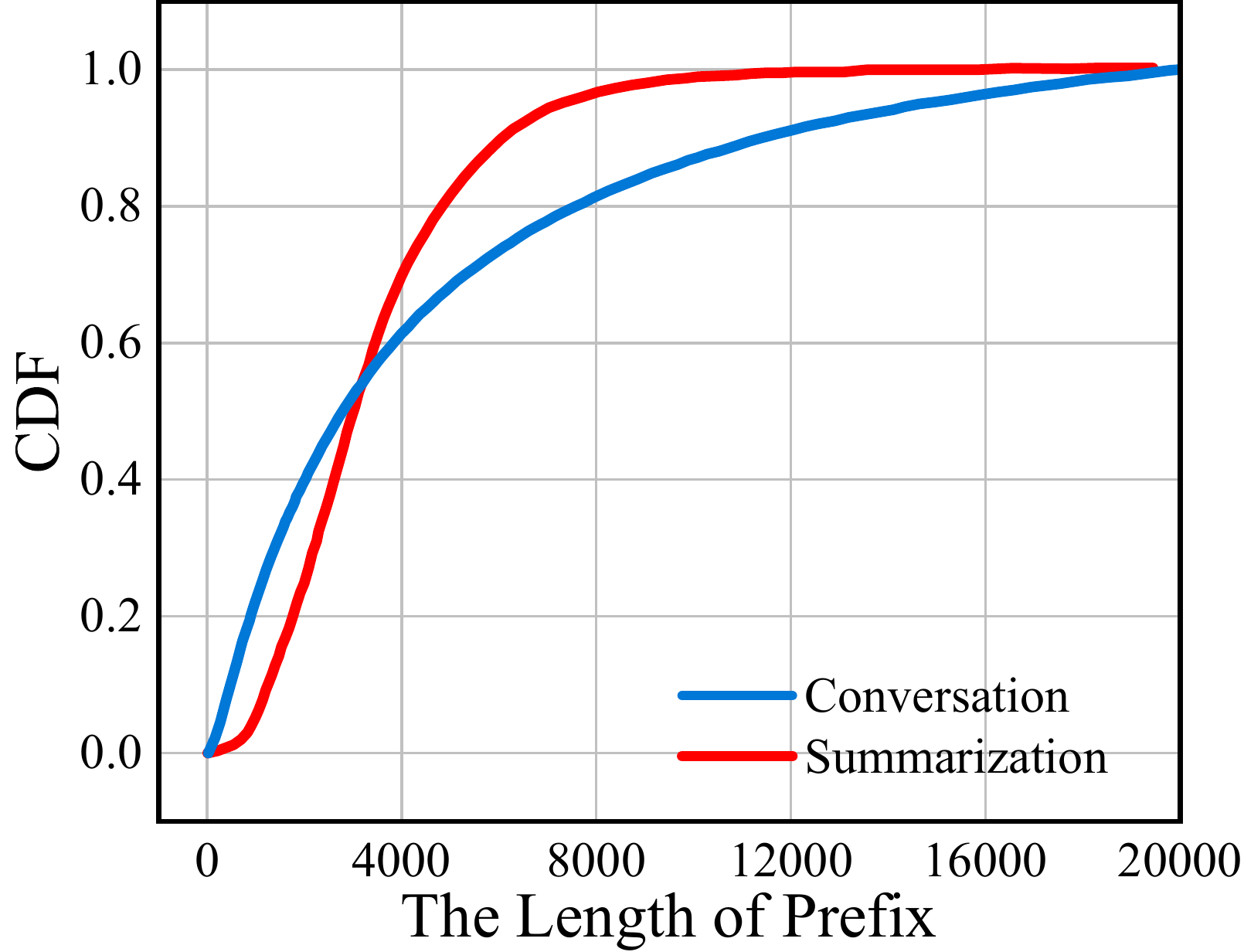} 
        \caption{GPT4 and summarization.}
    \end{subfigure}
    \caption{CDF of the lengths of prefix. } 
    \label{motivation-workload}
    
\end{figure}

Unfortunately, it is commonplace that tokens generated in the same iteration rely on different lengths of prefix for the following reasons. On one hand, the original input prompts are intrinsically of varied lengths. On the other hand, the number of tokens generated by each inference request differs greatly as well. And to the best of our knowledge, existing inference frameworks have not taken the length of prefix into account upon batching. They are likely to group requests with different lengths of prefix into a batch. To demonstrate the above conclusion, We analyze traces from different applications. The AzurePublicDataset~\cite{patel2024splitwise} published by Microsoft includes the two most widely used applications of LLM: \textit{conversation} and \textit{coding}. The Openchat\_ShareGPT4~\cite{wang2023openchat} includes conversations shared by users interacting with ChatGPT-4. The \textit{Summarization}~\cite{cohan-etal-2018-discourse} dataset is maintained by PubMed OpenAccess repository. Figure ~\ref{motivation-workload} presents the CDF (Cumulative Distribution Function) of the lengths of prefix involved in the inference for these traces. As shown in the figure, for the \textit{conversation}, about $5.66\%$ of generated tokens rely on the prefix with more than 2000 tokens; for the \textit{coding}, as much as $15.06\%$ of generated tokens rely on the prefix with even more than 4000 tokens. For the \textit{GPT4} and \textit{summarization}, the ratio of prefix longer than 4000 tokens can be as much as $40\%$. However, the results presented in Figure ~\ref{motivation1} argue that even the ratio of long prefix is as low as $6.25\%$ ($4/64$), the latency of an iteration will increase significantly. 

\subsection{Exploration of Prefix-aware Batching}

\begin{figure}[t]
    \centering
    \includegraphics[width=0.5\textwidth]{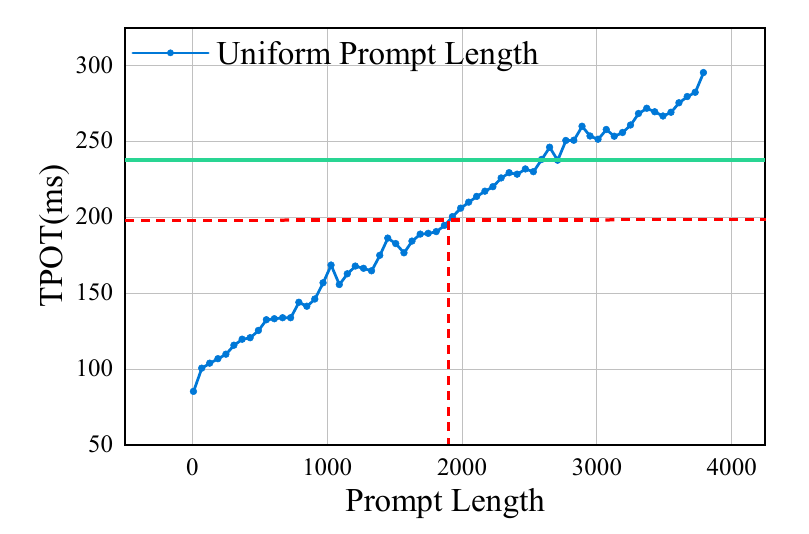}   
    \caption{Comparison of batching policies with and without considering the lengths of prefix.} 
    \label{motivation_group}
\end{figure}

As discussed above, a small fraction of long prefix will increase the latency of each iteration significantly. An intuitive optimization is grouping inference requests whose prefixes are of the similar length into a batch. To demonstrate the potential efficiency of this optimization, we conduct experiments to compare the batching policies with and without considering the lengths of prefix, respectively. In these experiments, we prepare 64 groups of inference prompts, where all prompts within a given group are of the same length, but prompts from different groups are of different lengths. Specifically, the lengths of prompts from the 64 groups are 10, 70, 130, 190,…, 3790 (the increment between two consecutive groups is 60), respectively. We feed these prompts to Llama2-7b on vLLM according to two different batching policies. The first policy takes each of the 64 groups as a batch, guaranteeing that prompts in a batch are of the same length. We run the 64 batches one by one and measure the average TPOT (Time per Output Token) for each batch. The blue line in Figure~\ref{motivation_group} presents the average TPOT of each of the 64 batches. As we can see, the longer input prompts introduce into a relatively higher TPOT. The horizontal red dotted line characterizes the average TPOT of all 64 batches, which is about $200ms$. The second batching policy selects one prompt from each of the 64 groups, generating a batch consisting of 64 prompts whose lengths are totally different. Apparently, the obtained 64 batches are the same, each of them contains 64 prompts with lengths of 10, 70, 130, 190,…, 3790, respectively. We run such a batch on Llama2-7b and measure the average TPOT as well. The horizontal green solid line in Figure~\ref{motivation_group} characterizes the average TPOT, which is about $233.43ms$.

In summary, we run the 4096 prompts on the same model (Llama2-7b) under two different batching policies. The prefix-aware batching policy guarantees that all tokens generated in a given iteration have the same cost, they are generated simultaneously and enter into the next iteration together. Whereas, the counterpart groups different lengths of prompts into a batch, some tokens relying on long prefix becomes the bottleneck, preventing other tokens from entering into the next iteration immediately even though these tokens have already been generated. During the period of waiting time, the computing resource (e.g., GPU) is underutilized. That is the reason why our prefix-aware batching policy outperforms the counterpart. Motivated by this observation, we propose a novel framework as well as the superior prefix-aware batching policy to support computing-efficient and high-throughput LLM inference.  

\section{AlignedServe}
In this section, we analyze the potential challenges introduced by the prefix-aware batching beforehand, and accordingly propose the overall architecture of AlignedServe as well as the prefix-aware batching and batch-level scheduling policies. The last subsection gives some details about the implementation and optimizations. 

\subsection{Challenges and Design Philosophy}   \label{subsection3.1} 

Grouping inference requests with the same length of prefix into a batch is non-trivial. The following challenges should be considered seriously and be overcome by well-designed architecture and algorithms. 

\textit{\textbf{Challenge 1: The serving system must maintain a very large amount of in-flight inference requests to guarantee that it is possible to select sufficient inference requests whose prefixes are of the similar length for a batch.}} As shown in Figure~\ref{motivation-workload}, the lengths of prefixes distribute in a very large range, from tens of tokens to more than $10,000$ tokens. In such a large range, it is unlikely to gather a batch of requests whose prefixes center around a small range, unless a very large amount of requests are waiting to be scheduled. However, maintaining so many in-flight requests introduces extremely high memory overhead, whereas GPUs are known to be memory-limited. To overcome this challenge, a straightforward solution is building a distributed serving system and offloading some data to the memory managed by CPUs. Accordingly, we propose a scalable architecture which splits prefill and decode across different GPUs, and dynamically swaps KVCache between GPU HBM and CPU memory on demand, as elaborated in Subsection~\ref{subsection3.2}.

\textit{\textbf{Challenge 2: The prefix-aware batching policy must dynamically adapt to workloads rather than simply grouping a fixed number (the batch size) of incoming inference requests into a batch.} }Traditional batching policies do not distinguish among inference requests. Once a sufficient number of incoming requests are accumulated, a new batch is generated and fed to the serving system. However, our prefix-aware batching policy not only cares about the number of incoming requests, but also considers the prefixes of these requests. Only when the serving system has accumulated a sufficient number of requests whose prefixes fall into a relatively small range, a satisfactory batch is generated. Unfortunately, these prefixes unevenly distribute in a very large range as shown in Figure~\ref{motivation-workload}. Our policy must dynamically adjust the window from which a batch of requests can be successfully selected. Furthermore, when a request departs from the batch due to completion, a new request should be scheduled to join the batch to improve computing efficiency. Our policy should also carefully choose an appropriate candidate request, rather than an arbitrary one. Accordingly, we propose an adaptive policy to greedily group incoming requests into batches, as elaborated in Subsection~\ref{subsection3.3}.

\textit{\textbf{Challenge 3: The serving system employing prefix-aware batching must be supported by a well-designed scheduler to achieve balance among throughput, latency, resource utilization and fairness.}} Traditional schedulers for LLM serving system mostly focus on the throughput and resource utilization, where both the two aspects do not conflict with each other since the improved resource utilization usually indicates much higher throughput. However, our prefix-aware batching policy does not feed an incoming request into the serving system immediately upon its arrival. The request may be required to wait for some other requests with the similar prefix. This optimization helps to improve the throughput and resource utilization, but at the cost of latency and fairness. Requests waiting for others to be grouped into the same batch are likely to endure longer latency. They may be even starved if there are no other requests that can be grouped together for a long period of time. To overcome the above drawbacks, we propose a smart scheduler which takes all the above considerations into account, as elaborated in Subsection~\ref{subsection3.4}.  

\subsection{The Overall Architecture}  \label{subsection3.2} 

\begin{figure}[t]
    \centering
    \includegraphics[width=0.7\textwidth, clip, trim=105pt 84pt 65pt 74pt]{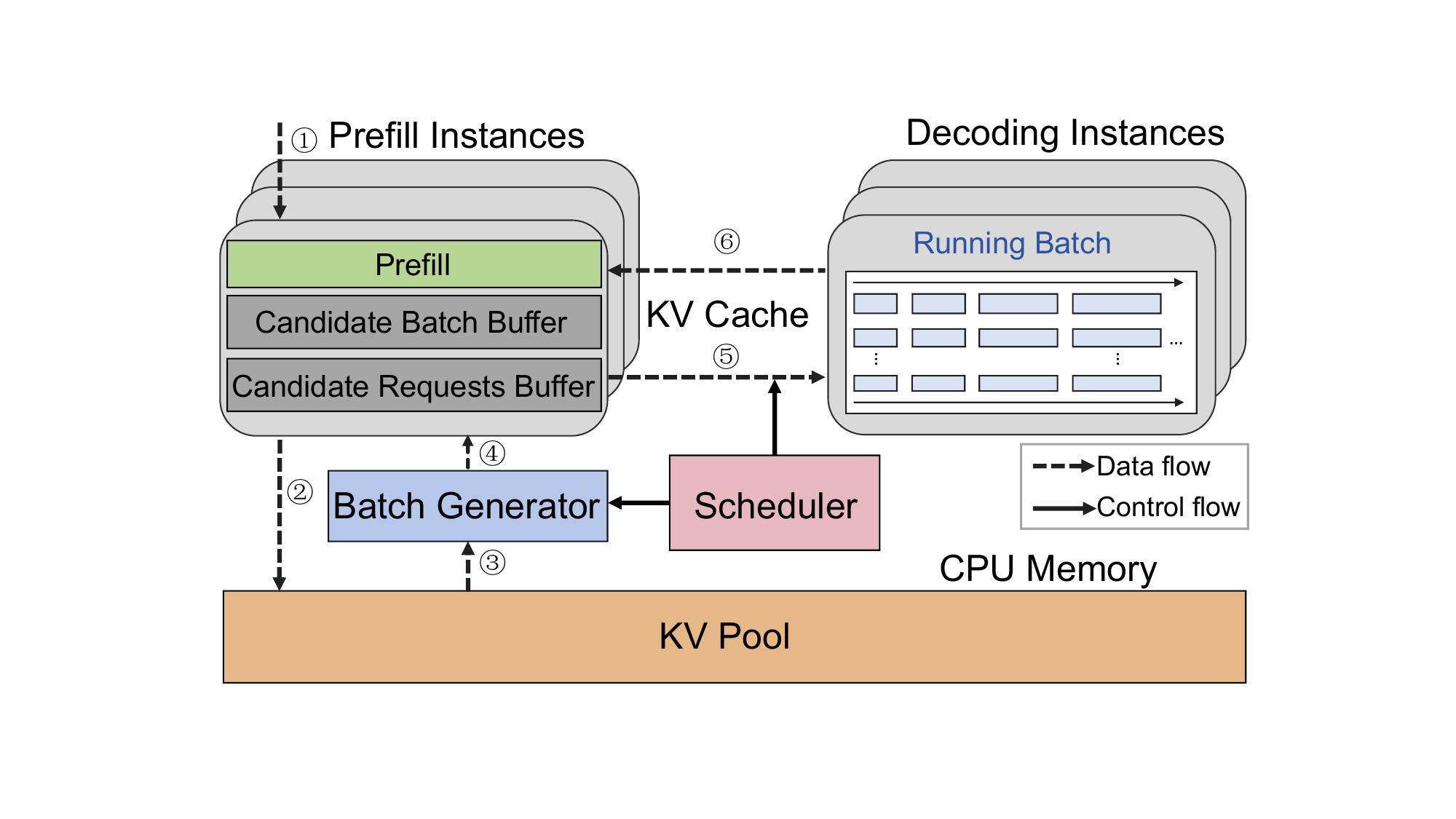}
    \caption{The overall architecture.}
    \label{fig:system_overview}
\end{figure}

As discussed in Subsection \ref{subsection3.1}, the prefix-aware batching requires a very large memory capacity to accommodate sufficient in-flight inference requests. To overcome this challenge, we propose a scalable architecture as shown in Figure~\ref{fig:system_overview}. The proposed serving system consists of three components, i.e., the KV pool, prefill instances and decoding instances, where both the prefill and decoding instances may be comprised of only one GPU, or multiple GPUs within the same server, or even many GPUs from different servers organized in a pipeline.

\textbf{KV pool:} The KV pool residing in host memory is used to keep the KVCache offloaded from GPU, guaranteeing that there are sufficient in-flight inference requests waiting to be batched. Compared with the limited HBM capacity within GPUs, the main memory managed by CPUs can be as large as several terabytes, which are able to accommodate the KVCache generated from millions of tokens (belonging to thousands of inference requests). Taking the Llama2-7B model as an example, the KV pair of a single token is approximately $512$KB, two terabytes of main memory is able to keep about 4 million tokens. Such a large number of tokens indicate that the serving system could ever successfully group a satisfactory batch. However, as demonstrated in Subsection \ref{subsection4.4}, the memory occupied by KV pool is no more than $250$GB under typical workloads.

\textbf{Decoding instances:} The decoding instances are responsible for accepting the scheduled batches and generating new tokens as in traditional serving systems. Note that the decoding instances may consist of multiple GPUs and can adopt any existing parallelism strategies such as data parallelism, tensor parallelism, and pipeline parallelism, while how to parallelize the decoding is beyond the scope of this work. We mostly focus on improving the computing efficiency of decoding instances.

\textbf{Prefill instances:} Traditionally, the prefill instances are simply responsible for processing the input prompts and initiating the KVCache for decoding instances. However in our framework, the prefill instances further act as the intermediate buffer for the KVCache swap between decoding instances and host memory. As shown in Figure~\ref{fig:system_overview}, the KVCache of in-flight requests are mostly kept in host memory. When a batch is scheduled to be run, a straightforward strategy is migrating the corresponding KVCache from host memory to the HBM of decoding instances directly via the PCIe interface between CPUs and GPUs. However, the bandwidth of PCIe is known to be limited. Instead, we propose to prefetch a prepared batch from host memory to the HBM of prefill instances, and further deliver the batch to decoding instances via NVLink between GPUs upon scheduling. To the best of our knowledge, this is the first work that employs one GPU to prefetch the KVCache for another by taking full advantage of the high bandwidth of NVLink. When the NVLink is unavailable, our framework gracefully falls back to the PCIe-only architecture, transmitting KVCache between CPU and GPU via PCIe directly. However, as NVLink has been widely adopted by high-end GPUs, and our work mostly focuses on the high-performance computing clusters, the novel GPU-Prefetch-For-GPU architecture works well in this kind of commonplace systems.

The three components described above are orchestrated by the batching and scheduling policies, as the data and control flows present in Figure~\ref{fig:system_overview}. When an inference request arrives, it is first processed by a prefill instance (Step \ding{172}), and the obtained KVCache is delivered to KV pool (Step \ding{173}). The batch generator selects appropriate requests from the large number of candidates in KV pool, and generates each batch by considering the length of prefixes (Step \ding{174}). The batch scheduled to be run in the near future is asynchronously prefetched to the Candidate Batch Buffer residing in prefill instances (Step \ding{175}), waiting to be forwarded to decoding instances via NVLink (Step \ding{176}). The scheduler monitors the decoding instances during runtime, and triggers scheduling in two cases. The first is that the running batch has generated an extremely large number of tokens, which exhaust the HBM capacity in decoding instances. In this circumstance, the scheduler evicts a request to Candidate Requests Buffer residing in prefill instances via NVLink (Step \ding{177}) by considering that the request will be rescheduled in the near future. The second case is that the running batch is unable to saturate the computing capability of decoding instances since the batch is too small. In this circumstance, the scheduler delivers some requests from Candidate Batch Buffer or Candidate Requests Buffer to decoding instances (Step \ding{176}). Once the Candidate Batch Buffer is empty, another batch is generated and prefetched to the buffer for the subsequent decoding. Note that the KVCache swapping between prefill and decoding instances is conducted via NVLink, whose high bandwidth helps to reduce the latency significantly, thereby eliminating the bubbles involved in batch scheduling.

\subsection{The Prefix-aware Batching Policy}  \label{subsection3.3}

As shown in Figure~\ref{fig:system_overview}, when the Running Batch is unable to saturate the computing capability of decoding instances, the scheduler will replace it with the Candidate Batch prepared in Candidate Batch Buffer via Step \ding{176} and \ding{177}. The Candidate Batch is produced by the Batch Generator before scheduling. Traditional Batch Generator mostly produces a batch by selecting inference requests according to the First-Come-First-Serve (FCFS) principle, where the inference requests with different lengths of prompts may be grouped into the same batch. But in this work, we aim to guarantee that the requests in a given batch have similar prefix lengths. To achieve this goal, we maintain a quad-tree to accommodate all the in-flight requests, as well as propose a Density First Search policy based on the tree to generate batches.

\begin{figure}[t]
    \centering
    \includegraphics[width=0.7\textwidth]{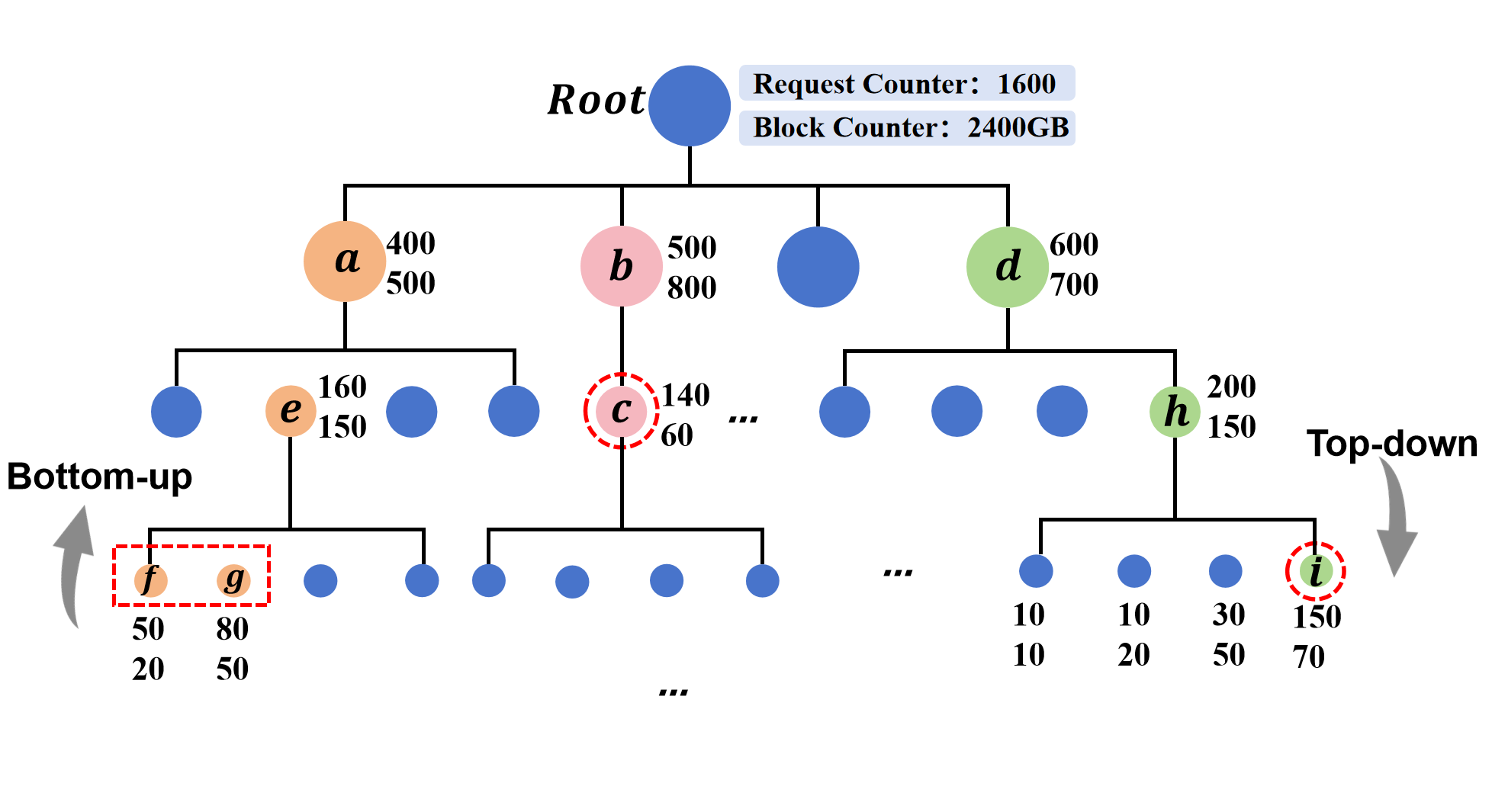} 
    \caption{Illustration of Density First Search, the two numbers associated with each internal node are the \textit{request counter} and \textit{block counter}, respectively.} 
    \label{quad_tree} 
\end{figure}

\textbf{The quad-tree structure.} As shown in Figure~\ref{quad_tree}, the quad-tree contains two types of nodes, i.e., the internal nodes and leaf nodes, where the internal nodes are used to guide the search, and the leaf nodes keep in-flight requests. Each internal node is responsible for a range of prefix lengths, which is further equally divided into four sub-ranges for its four child nodes. The iterative division of these ranges indicates that each internal node is the root of a sub-tree. We maintain a tuple (\textit{request counter}, \textit{block counter}) at each internal node to characterize the corresponding sub-tree, where the \textit{request counter} denotes the total number of in-flight requests kept in the sub-tree, and the \textit{block counter} denotes the total number of memory blocks consumed by the KVCache of these in-flight requests. Based on the quad-tree described above, the Density First Search policy aims to generate batches that are as large as possible while ensuring that requests within the same batch have similar prefix lengths. The quad-tree's larger branching factor reduces the tree height compared to binary trees, minimizing pointer chasing during Density First Search. Additionally, the node structure aligns well with CPU cachelines, improving the locality of memory accesses. We do not adopt the simple bucket-splitting strategy since this strategy imposes rigid boundaries that prevent the sliding window from consistently identifying optimal batches.

\textbf{Density First Search.} To generate a batch, the Density First Search policy starts the traverse from the root of quad-tree and conducts the top-down search. When arriving an internal node, e.g., \textit{a}, we check the \textit{request counter} and \textit{block counter} associated with \textit{a} to determine whether the requests kept in the corresponding sub-tree are able to be grouped into a batch. There are three cases to be considered upon the checking to an internal node.

\begin{algorithm}[t]
\caption{Density First Search Algorithm}
\label{alg:dynamic_batching}
\begin{algorithmic}[1]
\Input Maximum memory blocks $B_{\max}$, Minimum batch size $K_{\min}$
\Output Request in a batch $\mathcal{B}$
\Function{RecursiveDFS}{$node$, $B_{\max}$, $K_{\min}$}
   
   \State $\mathcal{B}$$\gets$ \Call{CollectRequests}{$node$},  $B_{\text{used}} $$\gets$ $\sum_{r \in \mathcal{B}} r.\textit{blocks}$

   \If{$B_{\text{used}} \leq B_{\max} \land |\mathcal{B}| \geq K_{\min}$}
       \Return $\mathcal{B}$\Comment{\textbf{Case 1}}
   \ElsIf{$B_{\text{used}} > B_{\max}$} \Comment{\textbf{Case 2}}
       \State \textbf{Find max density child $c_{\text{max}}$:}
       \State \Return \Call{RecursiveDFS}{$c_{\text{max}}$, $B_{\max}$, $K_{\min}$} 
   \Else \Comment{\textbf{Case 3}}
       \State $B_{\text{left}}$$\gets$$ B_{\max} - B_{\text{used}}$, $K_{\text{left}} $$\gets $$K_{\min} - |\mathcal{B}|$

       \If{$node.\textit{left\_sibling} \neq \emptyset$}
           \State $additon$$\gets$\Call{R-Search}{$node.\textit{left\_siblings}$,$B_{\text{left}}$,$K_{\text{left}}$}
       \Else
           \State $additon$$\gets$\Call{L-Search}{$node.\textit{right\_siblings}$,$B_{\text{left}}$,$K_{\text{left}}$}
       \EndIf
       \State $\mathcal{B}_{\text{final}}$$ \gets $$\mathcal{B} \cup additon[0:\min(K_{\text{left}}, |additon|)]$
       \State \Return $\mathcal{B}_{\text{final}}$

       \EndIf
\EndFunction

\Function{R-Search}{$siblings$, $B_{\text{left}}$, $K_{\text{left}}$}   
   \State $selected $$ \gets$$ \emptyset$
   \For{each $s$ in $siblings$ from right to left}
       \State $requests $$\gets$ \Call{CollectRequests}{$s$}
       \For{each $r$ in $requests$}
           \If{$r.\textit{blocks} + \sum(selected) \leq B_{\text{left}}$}
               \State $selected$$ \gets$ selected $\cup \{r\}$
            \Else
               \State \Return $selected$
           \EndIf
       \EndFor
   \EndFor
   \State \Return $selected$
\EndFunction

\Function{L-Search}{$siblings$, $B_{\text{left}}$, $K_{\text{left}}$}  
   \State $selected$ $\gets$ $\emptyset$
   \For{each $s$ in $siblings$ from left to right}
       \State $requests$ $\gets$ \Call{CollectRequests}{$s$}
       \For{each $r$ in $requests$}
           \If{$r.\textit{blocks} + \sum(selected) \leq B_{\text{left}}$}
               \State $selected$ $\gets$ $selected \cup \{r\}$
            \Else
                \State \Return $selected$
           \EndIf
       \EndFor
   \EndFor
   \State \Return $selected$
\EndFunction
\end{algorithmic}
\end{algorithm}

\begin{itemize}
    \item \textit{case 1: Generate a batch successfully.} If the total number of memory blocks (recorded in \textit{block counter}) occupied by the requests in the sub-tree does not exceed a given threshold, we think that these requests can be fully accommodated by the HBM of GPU, thus can be grouped into a batch. Taking the internal node \textit{c} as an example, the memory blocks are 60GB, which is less than the HBM capacity of an H100 GPU (80GB). In this circumstance, requests in the sub-tree (rooted at node \textit{c}) are grouped into a batch successfully. Note that, the prefixes of these requests range in the same sub-tree, thus are of the similar length. 

    \item \textit{case 2: Search top-down further.} If the total number of memory blocks recorded at the internal node \textit{d} exceeds the given threshold, indicating that the HBM of GPU cannot accommodate all the requests in the sub-tree according to \textit{case 1}, in this circumstance, we must conduct the top-down search further to shrink the search range. Specifically, among the four child nodes, the search comes to the one whose \textit{request counter} is larger than others. The largest \textit{request counter} indicates that the corresponding sub-tree has the highest request density. It is more likely to get a batch whose requests have similar prefix lengths in such a sub-tree. Taking the internal node \textit{h} as an example, among its four child nodes, the node \textit{i} has the largest \textit{request counter}, thus is selected as the search target in the next step.    

    \item \textit{case 3: Search bottom-up further.} If the total number of memory blocks does not exceed the given threshold mentioned in \textit{case 1}, but the total number of requests (recorded in \textit{request counter}) in the sub-tree is less than a threshold (e.g., 128), we do not group these requests into a batch even through they can be fully accommodated by the HBM of GPUs, since that the batch is too small to saturate the computing power of GPUs. Instead, we conduct the bottom-up search further to expand the search range. Taking the internal node \textit{g} as an example, its \textit{request counter} is only 80, which is too small to generate a batch. Accordingly, we return to its parent node \textit{e} to choose more requests from its left and/or right siblings (e.g., \textit{f} and \textit{g}).  
\end{itemize}

\textbf{Algorithm 1} presents the pseudo code of our Density First Search policy. As shown in the pseudo code, each generated batch is constrained by two input parameters, i.e., the total number of memory blocks occupied by the batch must be less than a threshold $B_{\max}$, and the total number of requests in the batch must more than a threshold $K_{\min}$. When the search arrives at a given node, we consider the three cases analyzed above, where \textit{case 1} corresponds to \textit{Line 4}, \textit{case 2} corresponds to \textit{Line 5}, and \textit{case 3} corresponds to \textit{Line 7}. For \textit{case 3}, the search must return to the parent node to expand the search range by further searching the left siblings (\textit{Line 10}) and/or right siblings (\textit{Line 12}). Note that the \textit{L-Search} function scans sibling nodes from right to left, whereas the \textit{R-Search} function scans sibling nodes from left to right. This mechanism guarantees that requests in the generated batch fall into as small a range as possible.

\subsection{The Batch-level Scheduling Policy}  \label{subsection3.4}


Traditional LLM serving systems mostly conduct the request-level scheduling, where all requests are scheduled independently. Taking the state-of-the-art continuous batching~\cite{orca} as an example, hundreds of requests are grouped into a batch, once a request in the running batch has finished, a new arriving request is added to the batch immediately, regardless of whether the arriving one is related to the running batch. On the contrary, we propose the batch-level scheduling policy, which schedules the incoming requests batch by batch, where the scheduled batches are generated by our prefix-aware batching policy proposed in Subsection~\ref{subsection3.3}.

To support the batch-level scheduling, we prepare two buffers at prefill instances as shown in Figure~\ref{fig:system_overview}, i.e., the Candidate Batch Buffer and the Candidate Requests Buffer, where the Candidate Batch Buffer keeps a batch waiting to be scheduled next, and the Candidate Requests Buffer keeps the remaining unscheduled requests belonging to the running batch. Unscheduled requests in the Candidate Request Buffer originate from two sources:
\begin{itemize}
    \item As the running batch iterates step by step, more and more tokens are generated, and the prefix of the running batch becomes longer gradually. In this period of time, some requests in the KV Pool may occasionally have similar prefix lengths to the running batch. Accordingly, we propose adding these requests to the running batch so that they can be decoded as soon as possible. Since these requests had not been grouped into the running batch by the prefix-aware batching policy, thus had not been kept in the Candidate Batch Buffer, we prefetch them to the Candidate Requests Buffer, where they wait for the chance to be decoded. 
    \item As the running batch generates more and more tokens, the produced KVCache will exhaust the HBM of GPUs. In this circumstance, we must evict some requests out of decoding instances to make space for the new generated tokens. When evicted from the running batch, these evicted requests are placed into the Candidate Request Buffer so that they have the chance to be rescheduled again in a short period of time.
\end{itemize}

Based on the framework described above, we elaborate the batch-level scheduling policy by taking a running batch as an example. The scheduling occurs when the running batch completes an iteration. Specifically, there are three distinct cases that warrant particular consideration.

\begin{figure}[t]
    \centering
    \begin{subfigure}[t]{0.45\linewidth}
        \centering
        \includegraphics[width=\textwidth, trim=0pt 330pt 470pt 0pt, clip]{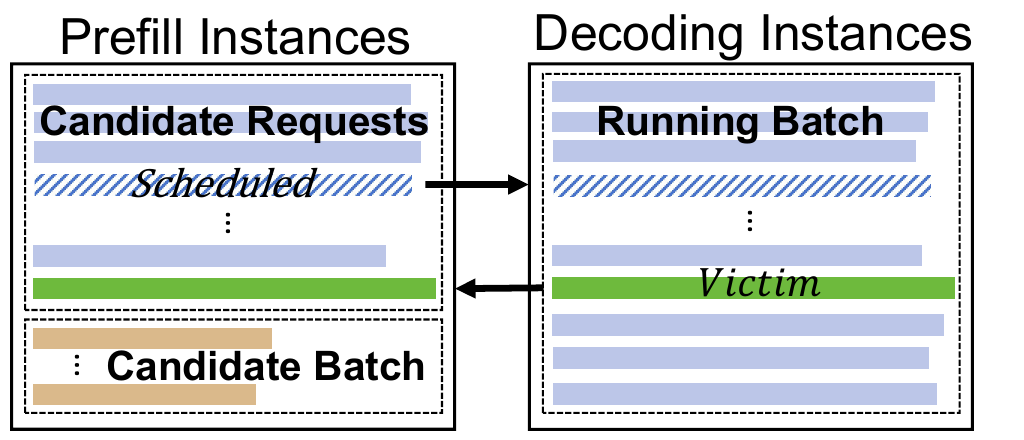}
        \caption{Case 1: Normal scheduling} 
        \label{subfig:case1} 
    \end{subfigure}%
    \begin{subfigure}[t]{0.45\linewidth}
        \centering
        \includegraphics[width=\textwidth, trim=0pt 330pt 470pt 0pt, clip]{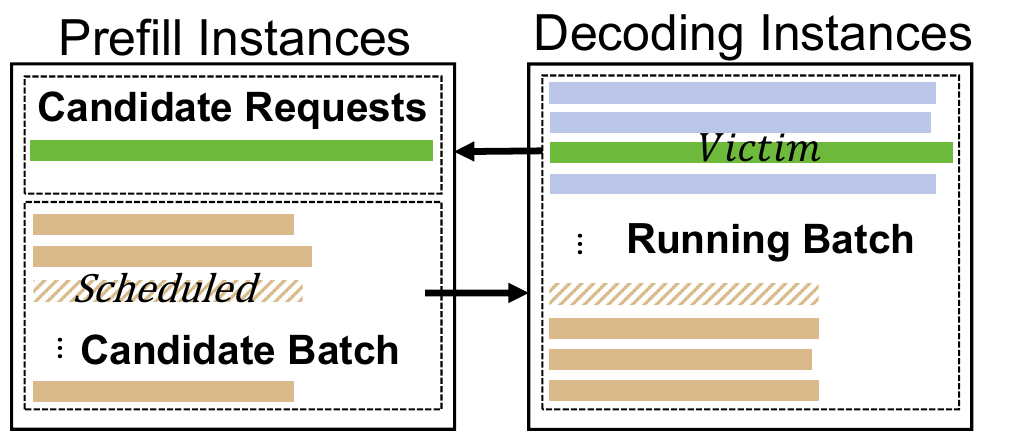}
        \caption{Case 2: Batch switch}
        \label{subfig:case2} 
    \end{subfigure}
    \caption{Illustration of (a) Scheduling from Candidate Requests Buffer, and (b) Scheduling from Candidate Batch Buffer}
    \label{fig:total_case} 
\end{figure}

\begin{itemize}
    \item \textit{case 1: Scheduling requests from Candidate Requests Buffer.} As Figure~\ref{subfig:case1} shows, when an iteration finishes, and some requests have completed in this iteration, the HBM capacity occupied by these completed requests will be released. Accordingly, we try to select some requests from the Candidate Requests Buffer to join the running batch. Note that requests in the Candidate Requests Buffer have the similar length of prefix with those in the running batch. Scheduling requests from Candidate Requests Buffer conforms to our primary design philosophy, i.e., serving requests with the similar length of prefix together. 

    \item \textit{case 2: Scheduling requests from Candidate Batch Buffer.} When an iteration finishes, and some requests have completed in this iteration, but there are no more requests existing in the Candidate Requests Buffer, indicating that all requests belonging to the running batch have been scheduled. In this circumstance, we try to select some requests from the Candidate Batch Buffer to join the running batch, as illustrated in Figure~\ref{subfig:case2}. Unfortunately, the selected requests are unlikely to have the similar length of prefix with those in the running batch. But we still add these requests to the running batch by considering that a larger batch is more likely to saturate the computing power of GPUs (on the contrary, if we refuse to add new requests to the decreasing running batch, the computing power of GPUs is underutilized). In this period of time, the running batch contains requests coming from two different original batches, and our framework behaves in the same manner as traditional continuous batching does. This is the very time when the scheduler switches one batch to another. Even though the process of batch switch violates our primary design philosophy (i.e., serving requests with the similar length of prefix together), as will be demonstrated in Subsection \ref{subsection4.4}, the negative impact introduced by batch switch is limited.

    \item \textit{case 3: Evicting requests to Candidate Requests Buffer.} When an iteration finishes, and there is not enough HBM space for keeping the tokens generated in the next iteration, some requests must be evicted from HBM to make space for the next iteration. In this circumstance, we must carefully select an appropriate victim from the running batch. Specifically, there are two scenarios that should be considered. The first is related to the \textit{case 1} described above, where all requests in the running batch have the similar prefix lengths. In this scenario, the longest request is selected to be evicted out, since the eviction of the longest request will release more HBM space. As illustrated in Figure~\ref{subfig:case1}, the green request is selected to be the victim. By considering that all requests in the running batch have the similar prefix lengths, we can also randomly select a request for simplicity. The second is related to the \textit{case 2} described above, where the requests in the running batch come from two different original batches (since a new batch is replacing the old one), accordingly, we select the longest victim from the old batch to evict the old batch out of HBM as soon as possible, thus accelerating the batch switch. As illustrated in Figure~\ref{subfig:case2}, the green request is selected to be the victim.   
    
\end{itemize}

\begin{algorithm}[t]
\caption{Batch-Level Scheduling Policy}
\label{alg:batch_level_scheduling}
\begin{algorithmic}[1]
\Input Running batch $B_{run}$, candidate requests buffer $R_{cand}$, candidate batch buffer $B_{cand}$
    \State \textit{completed\_reqs} $\gets$ Get\_Completed\_Requests($B_{run}$)
    \State Release\_HBM(\textit{completed\_reqs})
    \State Remove\_From\_Batch($B_{run}$, \textit{completed\_reqs})
    \If{not Has\_Sufficient\_HBM\_For\_Next\_Iteration()}
    \State\Comment{\textbf{Case 3}}
        \If{Is\_Batch\_Switching($B_{run}$)}
            \State \textit{victim} $\gets$ Select\_Victim\_From\_Old\_Batch($B_{run}$)
        \Else
            \State \textit{victim} $\gets$ Select\_Victim\_Longest($B_{run}$)  
        \EndIf
        \State Evict\_To\_Buffer($B_{run}$, \textit{victim}, $R_{cand}$)
    \Else
        \If{$R_{cand} \neq \emptyset$}
            \Comment{\textbf{Case 1}}
            \State \textit{new\_reqs} $\gets$ Select\_Requests\_From($R_{cand}$)
            \State Add\_To\_Batch($B_{run}$, \textit{new\_reqs})
        \ElsIf{$B_{cand} \neq \emptyset$}
            \Comment{\textbf{Case 2}}
            \State \textit{new\_reqs} $\gets$ Select\_Requests\_From($B_{cand}$)
            \State Add\_To\_Batch($B_{run}$, \textit{new\_reqs})
        \EndIf
    \EndIf
\end{algorithmic}
\end{algorithm}

\textbf{Algorithm 2} presents the pseudo code of our batch-level scheduling policy. The proposed scheduling policy presents superiority in two aspects. Firstly, each iteration can be executed rapidly since that the iteration-level bubbles have been mostly eliminated. Within an iteration, all the requests have the similar length of prefix, the costs of generating a token for each request are the same, there is no additional waiting delay introduced by the imbalanced workloads among requests. Secondly, the time interval between two consecutive iterations (which indicates system idle) can be significantly reduced. When an iteration completes, the scheduler may either add new requests to the running batch if some existing requests have finished, or remove requests from the running batch if the HBM lacks sufficient capacity for the next iteration. During this period, the iteration is paused awaiting scheduling. Traditionally, the scheduler transmits KVCache between GPU and CPU via PCIe interface, which is known to be very slow since the bandwidth of PCIe is limited. In our framework, the scheduler does not transmit KVCache via PCIe, but instead via the high performance NVLink. Specifically, when adding a request to the running batch, the scheduler transmits the corresponding KVCache from Candidate Requests Buffer or Candidate Batch Buffer (which reside in the HBM of prefill instances) to the HBM of decoding instances. When removing a request from the running batch, the scheduler transmits the corresponding KVCache from the HBM of decoding instances to the Candidate Requests Buffer. As the transmitting between GPUs is conducted by NVLink directly, the latency introduced by scheduling is reduced significantly. To the best of our knowledge, this is the first work that leverages prefill GPUs to prefetch or buffer KVCache for decode GPUs.

\subsection{Details of Implementation}  \label{subsection3.5}


Based on the overall framework described in the above subsections, we next highlight several critical implementation details that warrant careful consideration.\par
\textbf{Starvation:} 
Subsection~\ref{subsection3.3} presents our Density First Search algorithm which groups requests within a subtree into a batch. However, if there are no sufficient requests within a subtree, the subtree cannot generate a batch for a long time, accordingly requests within that subtree may suffer from starvation. To handle this situation, we set a time stamp for each internal node in the quad-tree, where the time stamp indicates how long the corresponding subtree has not generated a batch. If the time stamp associated with a subtree exceeds a given threshold, we give a higher priority to the subtree, thus avoiding potential starvation. The threshold can be dynamically adjusted to achieve different service-level objectives(SLOs).

\textbf{Dynamic scheduling:} In Subsection~\ref{subsection3.3}, the prefix-aware batching policy statically generates batches for scheduling. The generated batches are prefetched into the Candidate Batch Buffer to wait for being scheduled. Once a batch has been scheduled to be run, the decoding proceeds constantly, and some requests in the KV pool may occasionally have the similar length of prefix with those in the running batch. In this circumstance, the scheduler dynamically prefetches these requests into the Candidate Requests Buffer, where they wait for being added to the running batch. This optimization also helps to avoid starvation, since some requests failing to be grouped into a static batch may have the chance to be dynamically scheduled. 

\textbf{Prototype:} We implement our framework based on the state-of-the-art DistServe~\cite{zhong2024distserve}. DistServe is built upon a novel serving architecture that disaggregates the prefill and decode stages of LLM inference. After the prefill stage, requests are immediately transmitted to decoding instances for subsequent token generation, where all the requests are managed by the FCFS scheduling policy. Our AlignedServe also adopts the disaggregated architecture. We improve DistServe by incorporating a prefix-aware batching and a batch-level scheduling policy, obtaining our prototype system. Specifically, when a request has been processed by the prefill stage, it will not be transmitted to the decode stage immediately as DistServe does, but be kept in the KV Pool instead. Requests in the KV Pool are grouped by the prefix-aware batching policy, and the generated batches will be scheduled by the batch-level scheduling policy ultimately. Compared to DistServe, AlignedServe introduces additional latency, especially when a large number of requests in prefill instances already have similar prompt lengths. However, as shown in Figure~\ref{motivation-workload}, the lengths of prefixes fall into a very large range, thus it is unlikely to accumulate enough requests whose prompts are of similar length in the limited HBM of prefill instances. Furthermore, batching in the KV pool does introduce additional latency, but helps to significantly improve the throughput, which attracts much more attention. So, our optimization remains meaningful. 


\section{Performance Evaluation}

This section evaluates our AlignedServe via experiments driven by both synthetic and application workloads. The comparison candidates include the state-of-the-art vLLM~\cite{vllm2024}, DistServe~\cite{zhong2024distserve}, and DeepSpeed-FastGen~\cite{holmes2024deepspeed}. For brevity, we refer to DeepSpeed-FastGen as FastGen in the remainder of this paper. We conduct experiments to demonstrate that AlignedServe outperforms the others in terms of \textit{Throughput} and \textit{P99 Latency} about decoding.

\subsection{Experimental Setup}

\textbf{System configuration and models.} We run experiments on a server equipped with 2 Intel(R) Xeon(R) Platinum 8462Y+ CPUs, 8 NVIDIA H100 GPUs connected with NVLink, and 800GB DRAM. The GPUs and CPUs are connected with PCIe 5.0. Similar with the previous LLM inference serving system~\cite{zhong2024distserve, IMPRESS}, we use the OPT models~\cite{DBLP:journals/corr/abs-2205-01068} to evaluate our framework, where the models cover OPT-2.7B, OPT-6.7B, OPT-13B and OPT-30B, we use FP16 precision in all experiments. Some important parameters are set as follows. The whole range of prefix lengths managed by the quad-tree is set to [1,65536]. The lengths longer than 65536 are all considered to be 65536 for simplicity. $B_{max}$ in \textbf{Algorithm 1} is specifically set to 40\% of total GPU blocks, and $K_{min}$ is set to 36. These parameters are configurable based on hardware specifications and workloads.

\textbf{Workloads.} Our AlignedServe is evaluated by both synthetic benchmarks and application workloads. About the synthetic benchmarks, we have constructed several datasets which contain different ratios of long requests and short requests, where the number of tokens contained by a short request is set to be less than one thousand, and the number of tokens contained by a long request ranges from one thousand and eight thousand. About the application workloads, we adopt the well-known ShareGPT, LongBench and AzurePublicDataset. The ShareGPT dataset consists of publicly shared multi-round conversational interactions between users and LLMs. LongBench dataset is a comprehensive bilingual benchmark test designed to evaluate the capabilities of LLMs for long context understanding.  AzurePublicDataset offers real-world inference traces, including request patterns and input/output lengths, making it suitable for evaluating production-like serving behavior.

\textbf{Baseline.} We compare AlignedServe with the state-of-the-art works including vLLM~\cite{vllm2024}, DistServe~\cite{zhong2024distserve}, and FastGen ~\cite{holmes2024deepspeed}. vLLM integrates the technologies proposed by SarathiServe and Orca, serving the inference requests in the FCFS manner. DistServe is a popular prefill/decode–disaggregated architecture, which is adopted by AlignedServe as well. FastGen is designed for high-throughput LLM inference, which is also pursued by AlignedServe. About some other state-of-the-art works, LoongServe targets to extremely long-sequence serving and focuses on optimizing parallel strategies, which is orthogonal to our approach. HotPrefix, Preble, and BatchLLM focus on the shared prefixes across different requests to eliminate redundant computation, rather than addressing the performance degradation caused by varied prefix lengths. So, we need not to compare AlignedServe with them.

\textbf{Metrics.} The performance is measured by \textit{Decoding Throughput} and \textit{P99 Latency of TPOT}(Time Per Output Token). Both the two metrics mostly focus on decoding since the output of decoding is the straightforward performance concerned by users. Besides the comparison in terms of performance, we analyze the overhead and ablation further via experiments to give a much more deeper understanding to our framework, explaining why our AlignedServe outperforms the others.

\subsection{Decoding Throughput}  \label{section4.2}

\begin{figure*}[t] 
    \centering 
    
    \begin{subfigure}{\textwidth}
        \centering
        \includegraphics[width=0.8\textwidth]{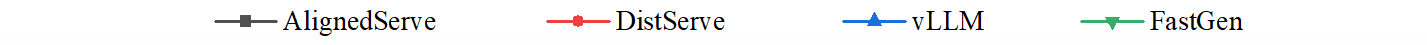}
    \end{subfigure}
    
    \begin{subfigure}[t]{0.248\textwidth}
        \centering
        \includegraphics[width=\textwidth]{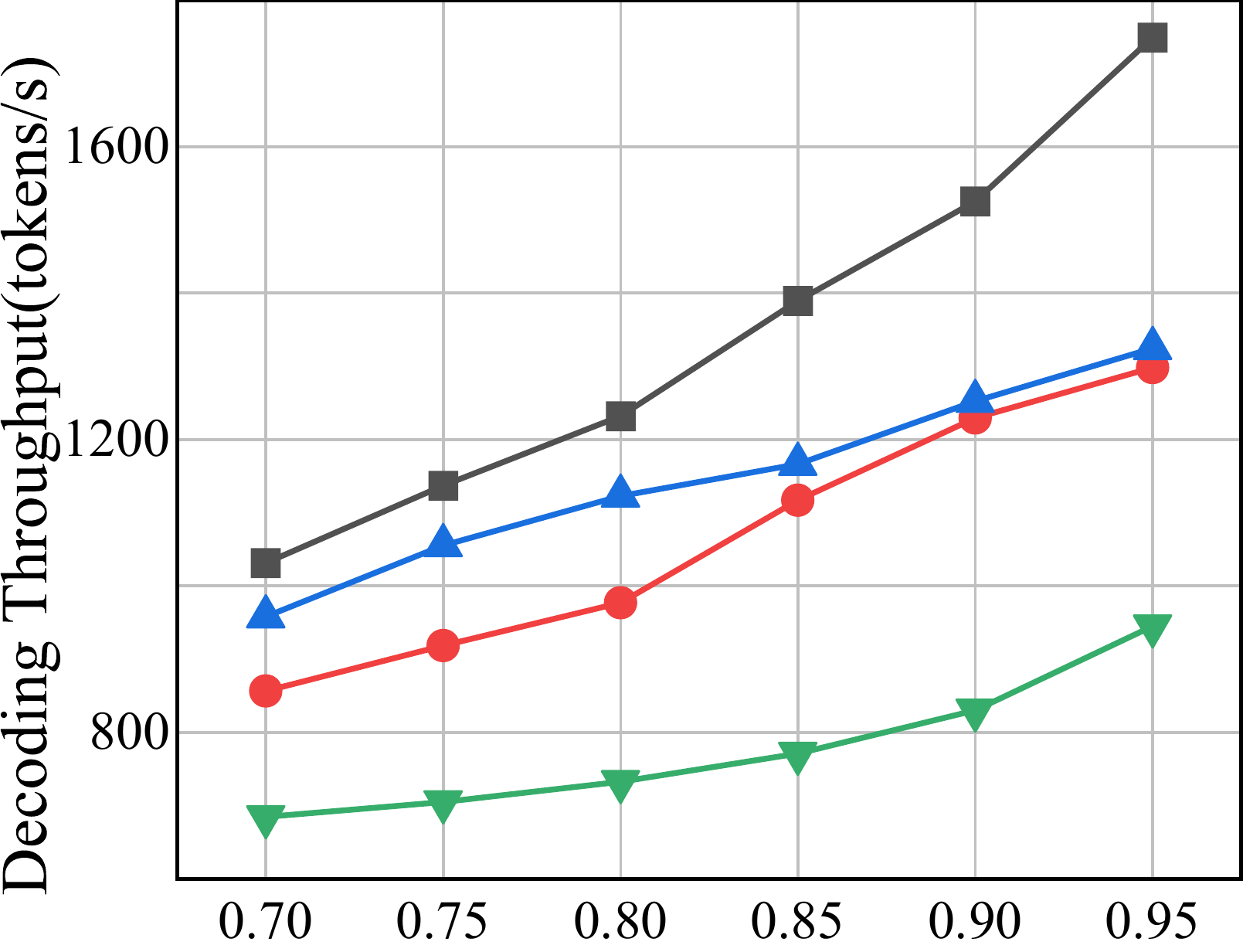}
        \caption{OPT-2.7B}  
        \label{fig:opt27b}   
    \end{subfigure}%
    \begin{subfigure}[t]{0.248\textwidth}
        \centering
        \includegraphics[width=\textwidth]{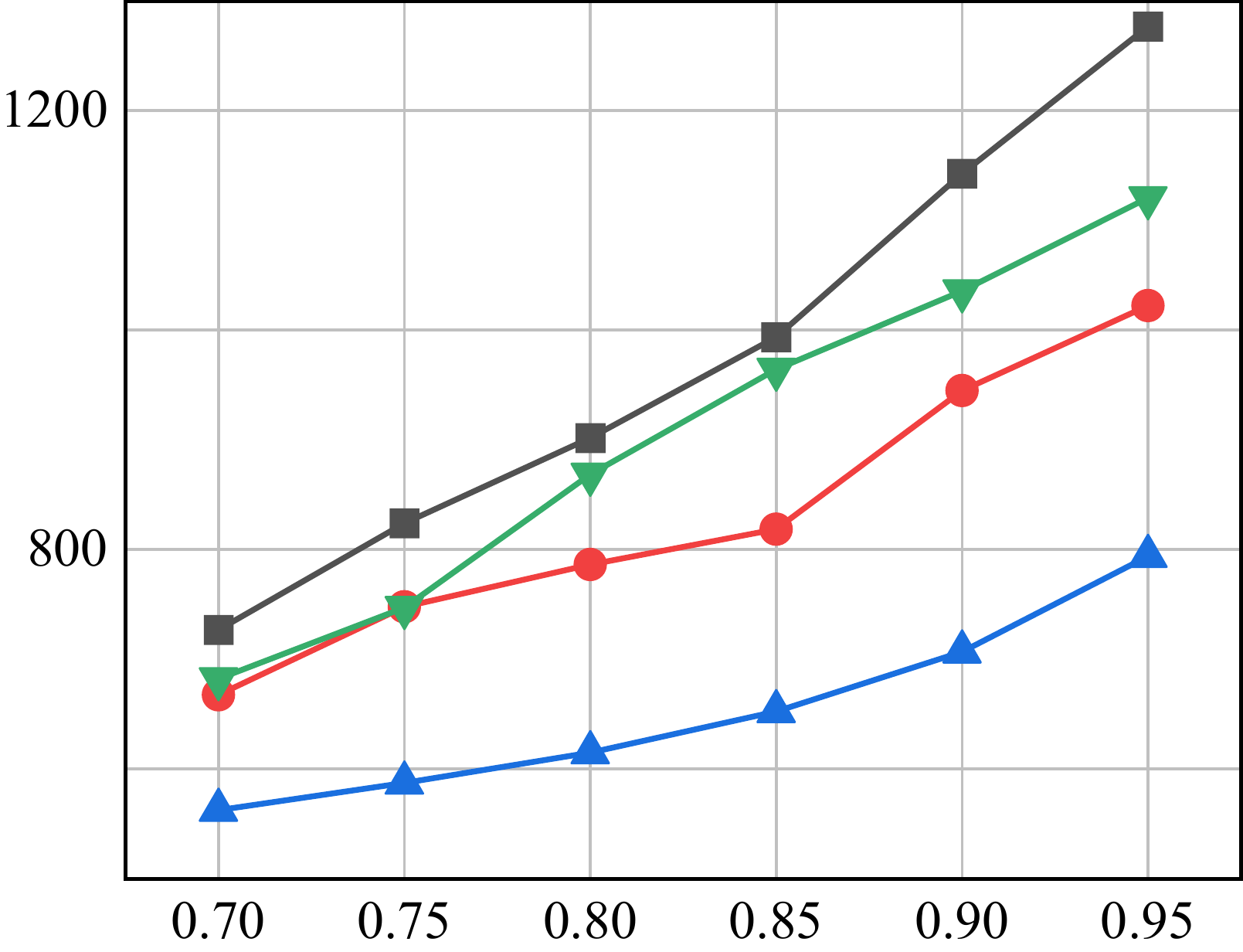}
        \caption{OPT-6.7B}
        \label{fig:opt67b}
    \end{subfigure}
    \begin{subfigure}[t]{0.248\textwidth}
        \centering
        \includegraphics[width=\textwidth]{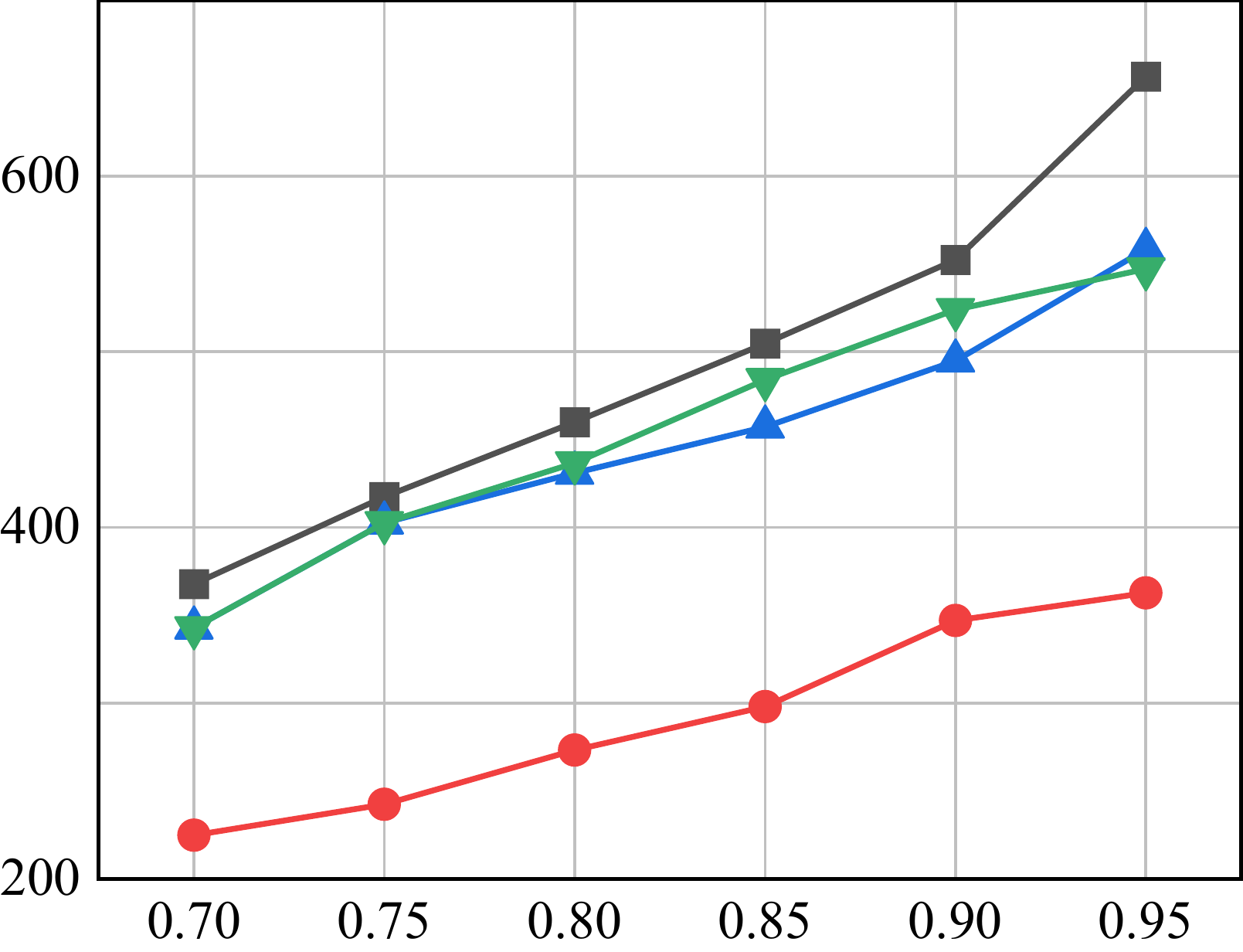}
        \caption{OPT-13B}
        \label{fig:opt13b}
    \end{subfigure}%
    \begin{subfigure}[t]{0.248\textwidth}
        \centering
        \includegraphics[width=\textwidth]{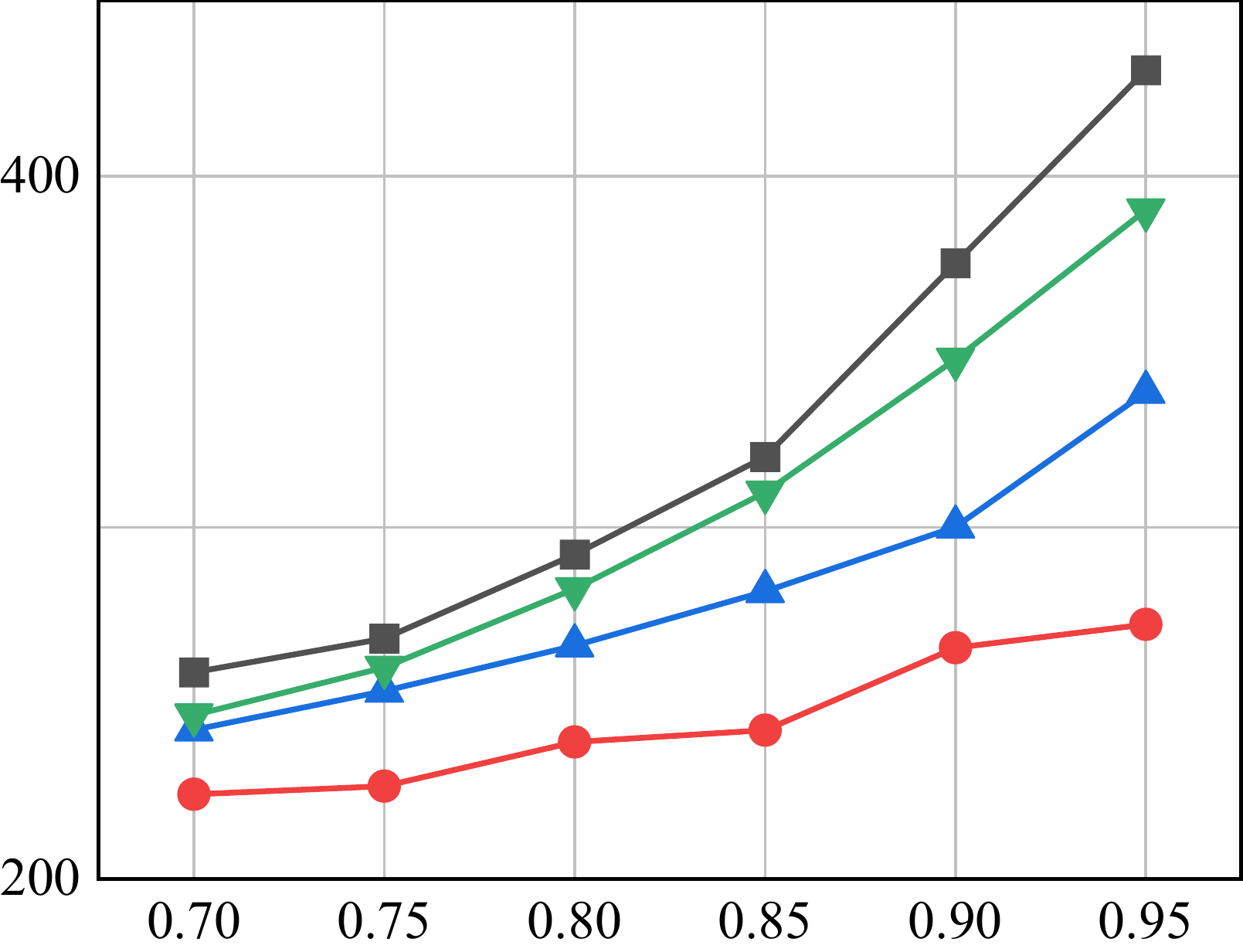}
        \caption{OPT-30B}
        \label{fig:opt30b}
    \end{subfigure}
    
    \caption{Decoding throughput (tokens/s) on synthetic workloads.}
    \label{evaluation_decode_throughput}
\end{figure*}

\begin{figure}[t]
    \begin{subfigure}[t]{0.3\linewidth}
        \centering
        \includegraphics[width=\textwidth]{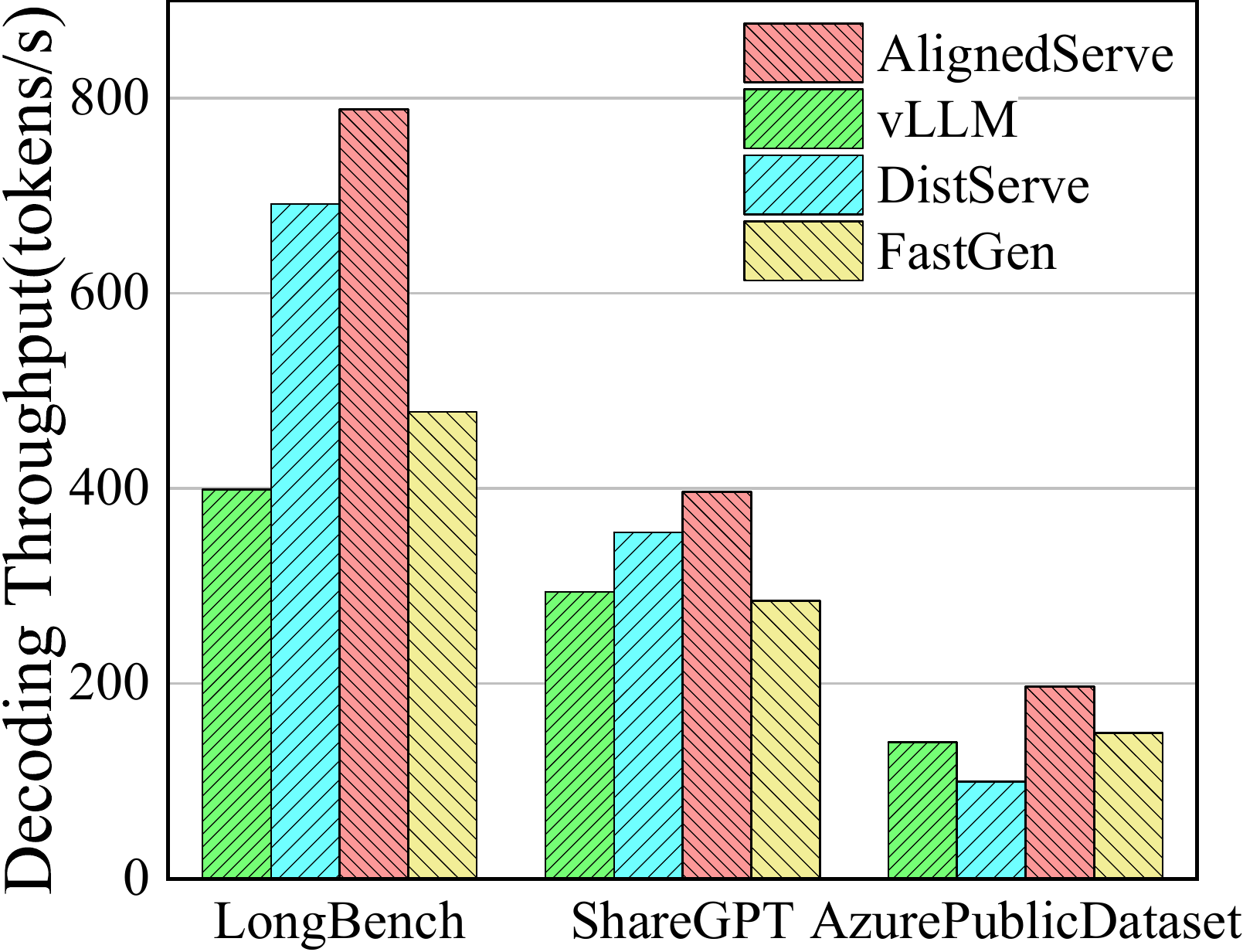}
        \caption{OPT-6.7B.}
    \end{subfigure}
    \begin{subfigure}[t]{0.3\linewidth}
        \centering
        \includegraphics[width=\textwidth]{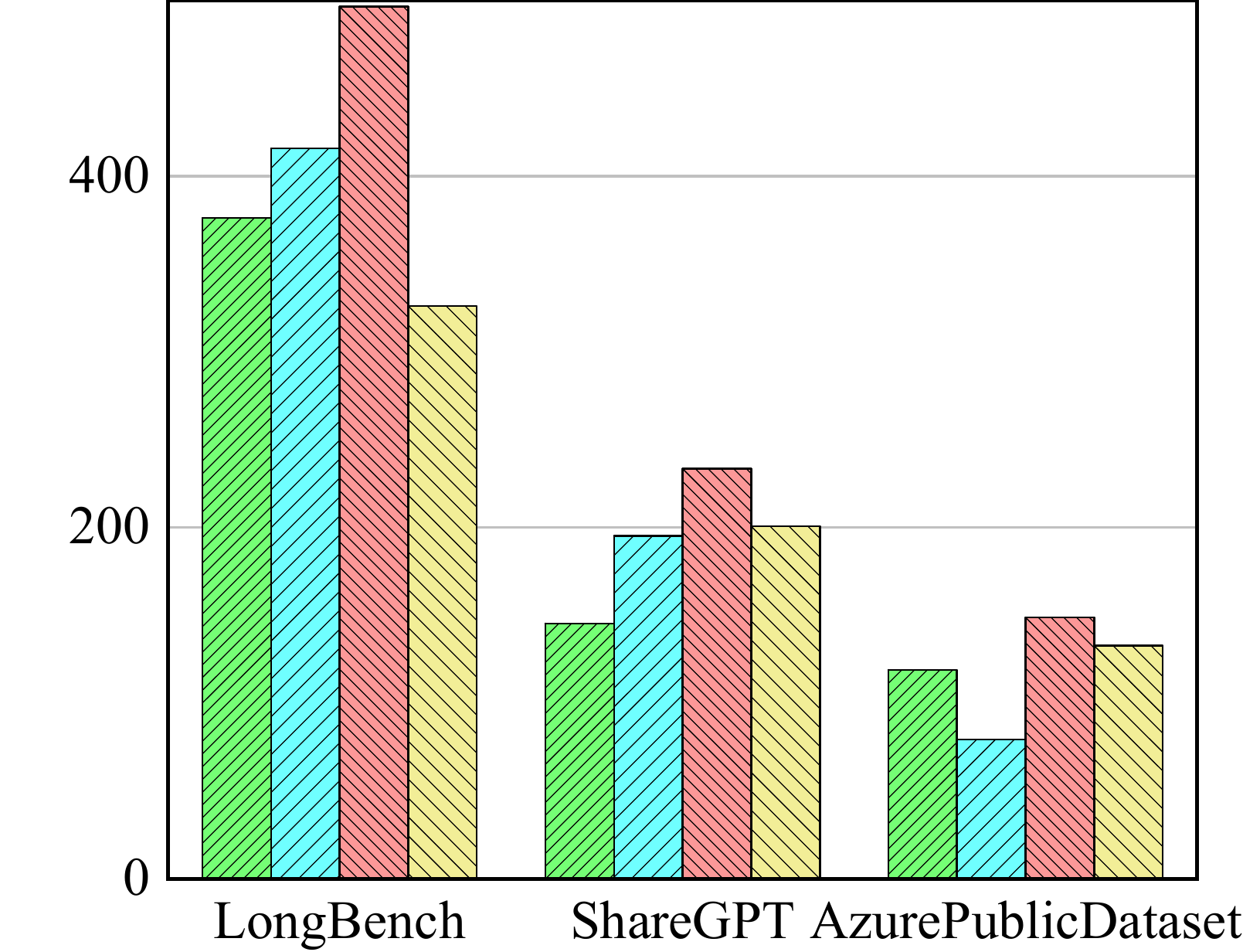}
        \caption{OPT-13B.}
    \end{subfigure}
    \caption{Decoding Throughput on application workloads.} 
    \label{open_dataset_throuhghput}
    \vspace{-15pt}
\end{figure}

The major principal behind our work is reducing the negative impact introduced by these long requests. So, the first experiment is designed to evaluate the comparison candidates under workloads with different ratios of long requests and short requests. Specifically, we generate six workloads, where the ratios of short requests in these workloads are 70\%, 75\%, 80\%, 85\%, 90\%, 95\%, respectively. The performance is measured by \textit{Decoding Throughput}, which has also been adopted in many state-of-the-art works~\cite{10.5555/3618408.3619696,TightLLM}. Experimental results are presented in Figure~\ref{evaluation_decode_throughput}, where the horizontal axis denotes different ratios of short requests. As shown in the figure, our AlignedServe outperforms the others constantly under all workloads. Especially, when the short requests account for 95\% of the total number of requests (thus the long requests account for the left 5\%), AlignedServe presents significant superiority. For example, on the OPT-2.7b model, AlignedServe outperforms vLLM, DistServe and FastGen by 1.32$\times$, 1.35$\times$, and 1.85$\times$, respectively. According to the Pareto principle (80/20 rule), mixing a small fraction of long requests with a large amount of short requests is commonplace, while even this small fraction of long requests will degrade the performance remarkably, as demonstrated in Subsection~\ref{section2.3}. We propose the prefix-aware batching policy to overcome this challenge, thus improve the performance significantly.

We further demonstrate the effectiveness of our proposal by three publicly released application workloads collected from production systems, i.e., ShareGPT, LongBench and AzurePublicDataset. The experimental results are presented in Figure~\ref{open_dataset_throuhghput}. As shown in the figure, our AlignedServe outperforms the others by a maximum of 1.98$\times$, 1.35$\times$ and 1.98$\times$ upon LongBench, ShareGPT and AzurePublicDataset, respectively, indicating that AlignedServe behaves well on realistic applications. In Subsection~\ref{section2.3}, Figure~\ref{motivation-workload} reveals that the length of requests in AzurePublicDataset falls into a very large range, from 3 to 7437. This observation indicates a much more significant negative impact introduced by these long requests. The comparison candidates such as vLLM and DistServe suffer from this negative impact, but on the contrary our framework is able to handle this situation efficiently and therefore achieves higher performance.

\subsection{Decoding Latency}   \label{section4.3}

\begin{figure*}[t] 
    \centering 
    \begin{subfigure}{\textwidth}
        \centering
        \includegraphics[width=0.6\textwidth]{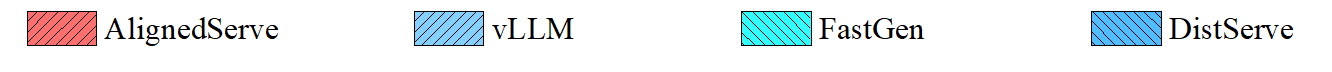}
    \end{subfigure}
    
    \begin{subfigure}[t]{0.248\textwidth} 
        \centering
        \includegraphics[width=\textwidth]{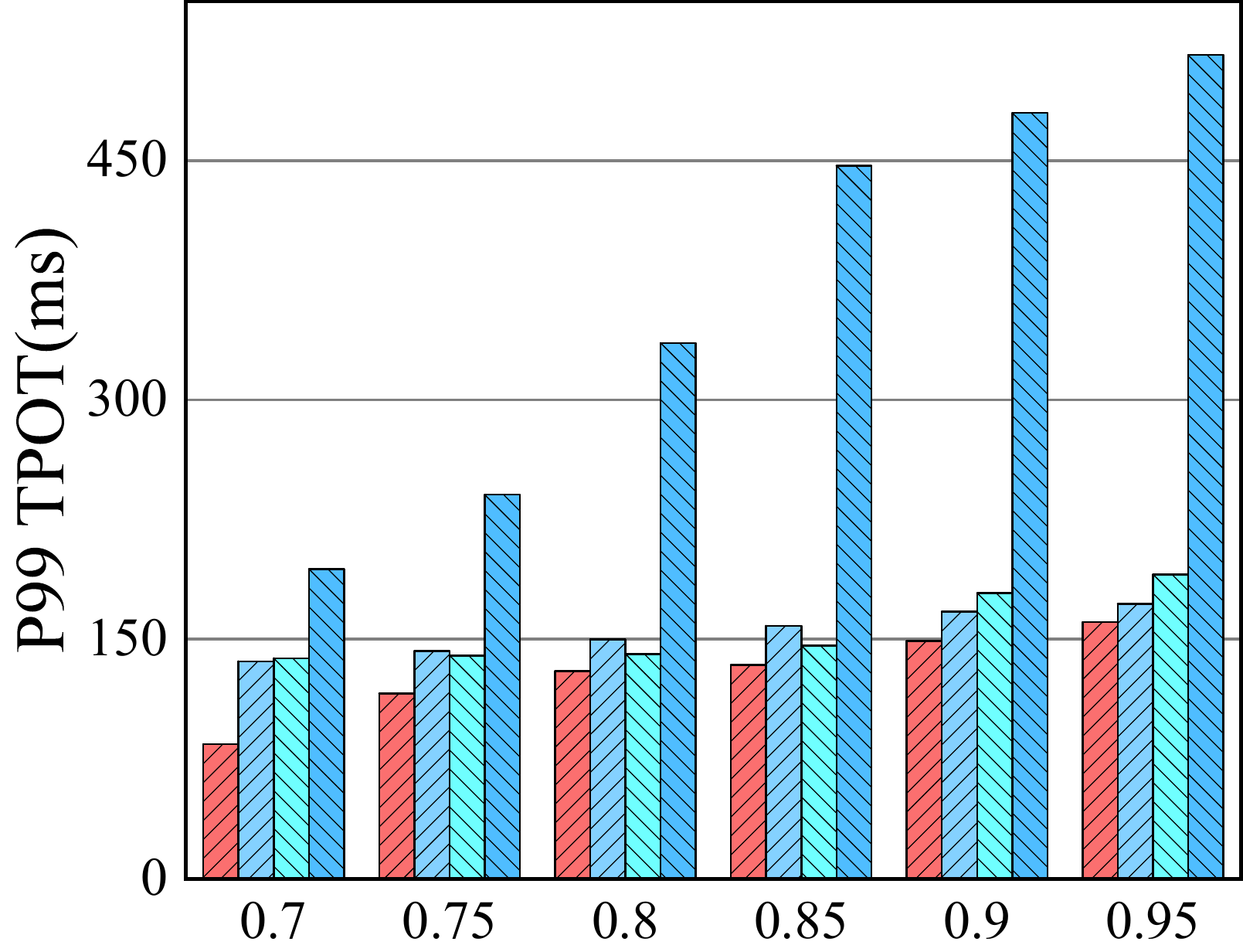}
        \caption{OPT-2.7B.}
    \end{subfigure}%
    \begin{subfigure}[t]{0.248\textwidth}
        \centering
        \includegraphics[width=\textwidth]{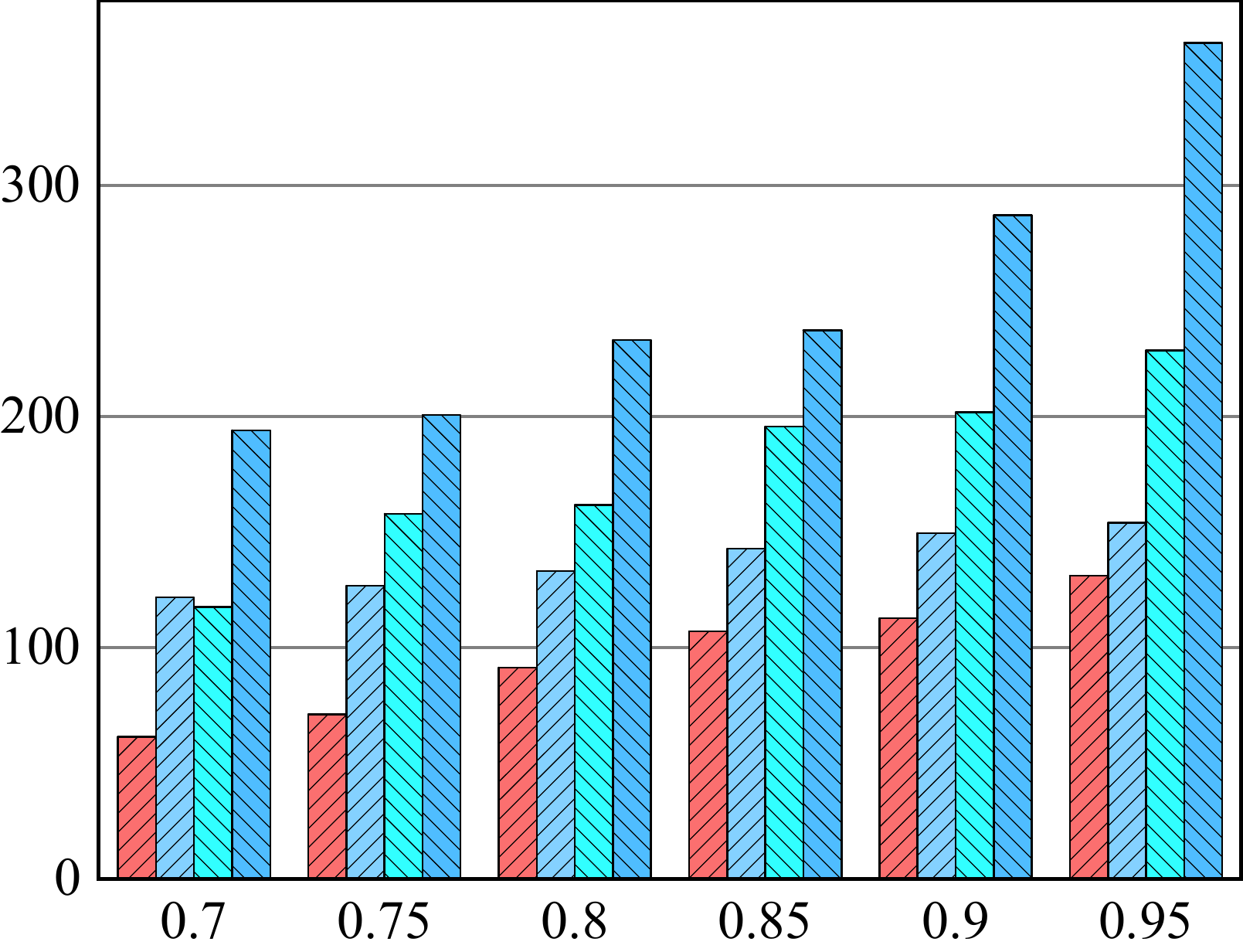}
        \caption{OPT-6.7B.}
    \end{subfigure}
    \begin{subfigure}[t]{0.248\textwidth}
        \centering
        \includegraphics[width=\textwidth]{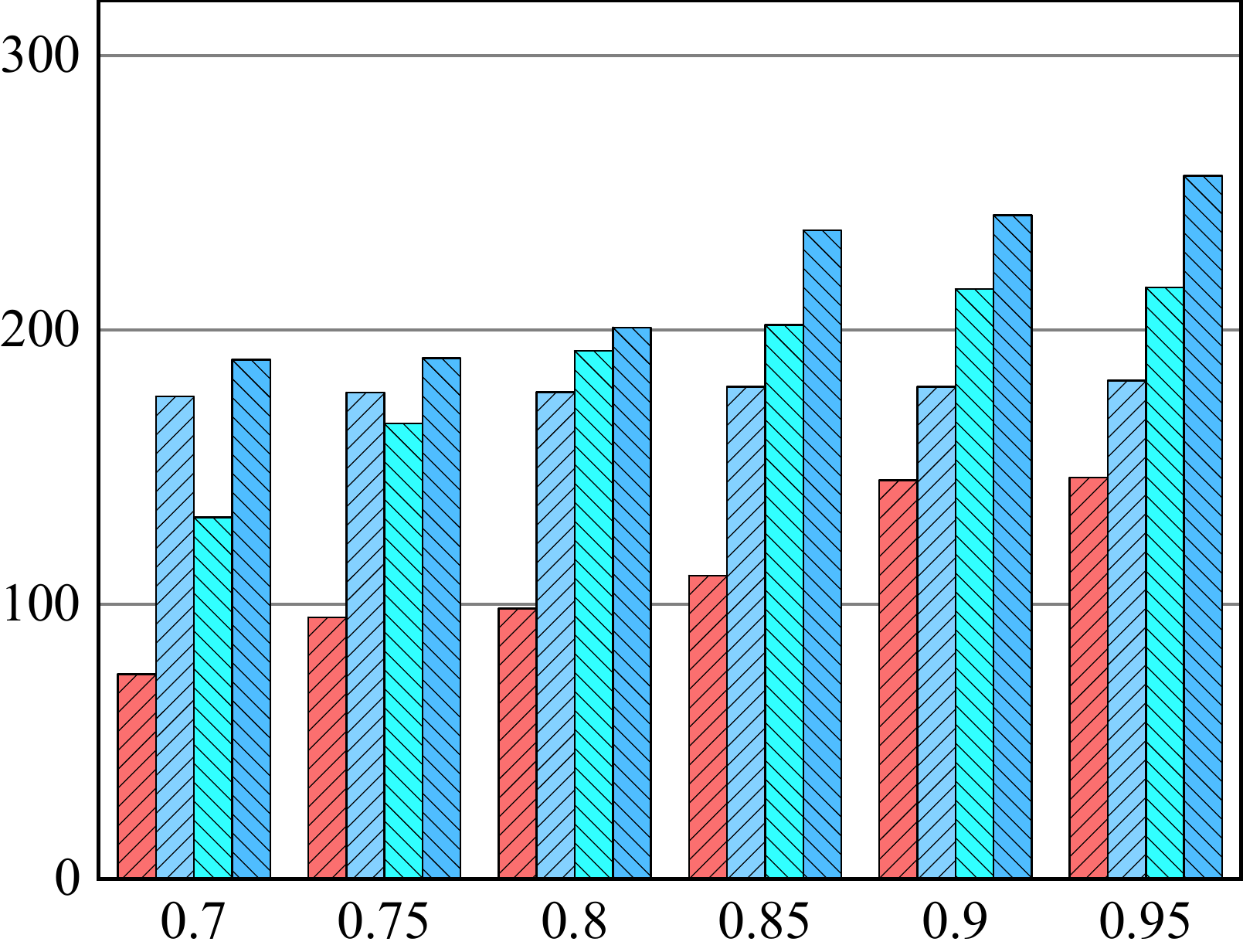}
        \caption{OPT-13B.}
    \end{subfigure}%
    \begin{subfigure}[t]{0.248\textwidth}
        \centering
        \includegraphics[width=\textwidth]{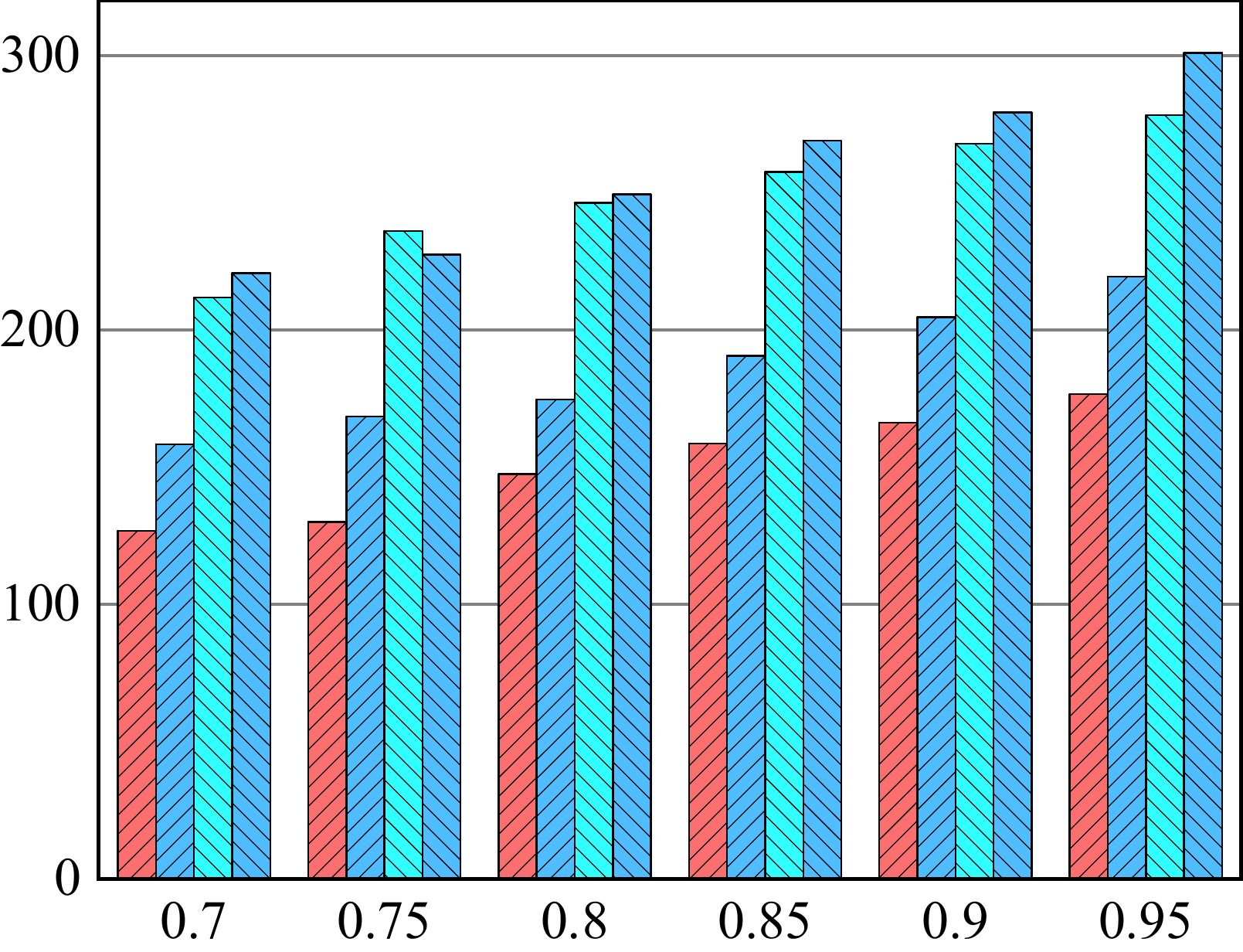}
        \caption{OPT-30B.}
    \end{subfigure}
    \caption{P99 TPOT on synthetic workloads.}
    \label{evaluaton_tpot}
\end{figure*}

As analyzed in Subsection~\ref{section2.3}, some long requests within a batch introduce iteration-level bubbles, which indicate the increased latency of each iteration. We pursue to eliminate the iteration-level bubbles by adopting the prefix-aware batching and batch-level scheduling policy, thus endeavor to reduce the latency. To demonstrate the achievement of our proposal, we conduct experiments to evaluate the \textit{P99 Latency of TPOT}, which characterizes the latency of each decoding iteration. The synthetic workloads are generated in the same manner as Subsection~\ref{section4.2} has done. Experimental results are shown in Figure~\ref{evaluaton_tpot}. As the figure indicates, our AlignedServe consistently achieves substantially lower TPOT than vLLM, DistServe, and FastGen across different models and workloads, yielding 1.74$\times$-3.05$\times$ lower latency.

\begin{figure}[t]
    \begin{subfigure}[t]{0.3\linewidth}
        \centering
        \includegraphics[width=\textwidth]{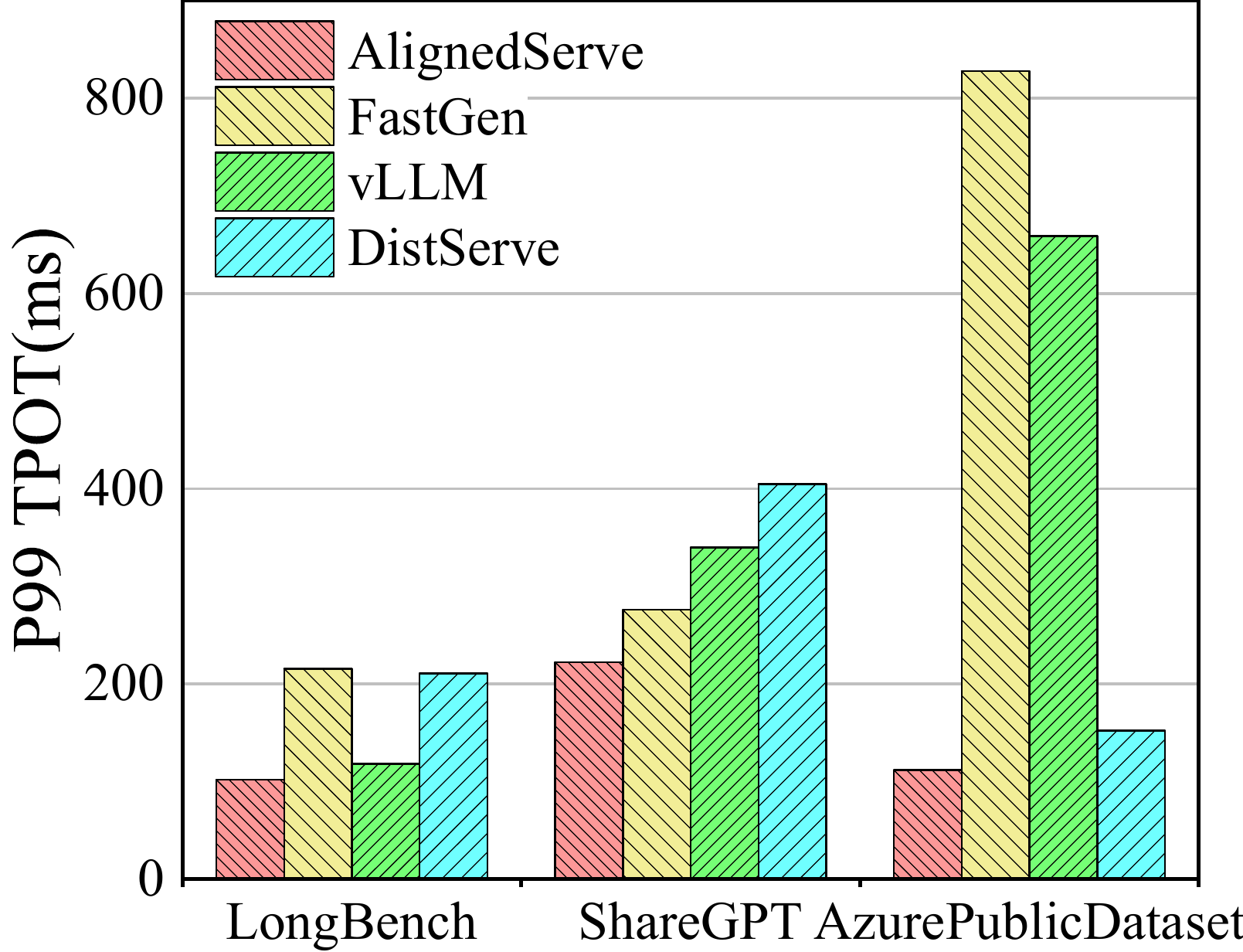}
        \caption{OPT-6.7B.}
    \end{subfigure}%
    \begin{subfigure}[t]{0.3\linewidth}
        \centering
        \includegraphics[width=\textwidth]{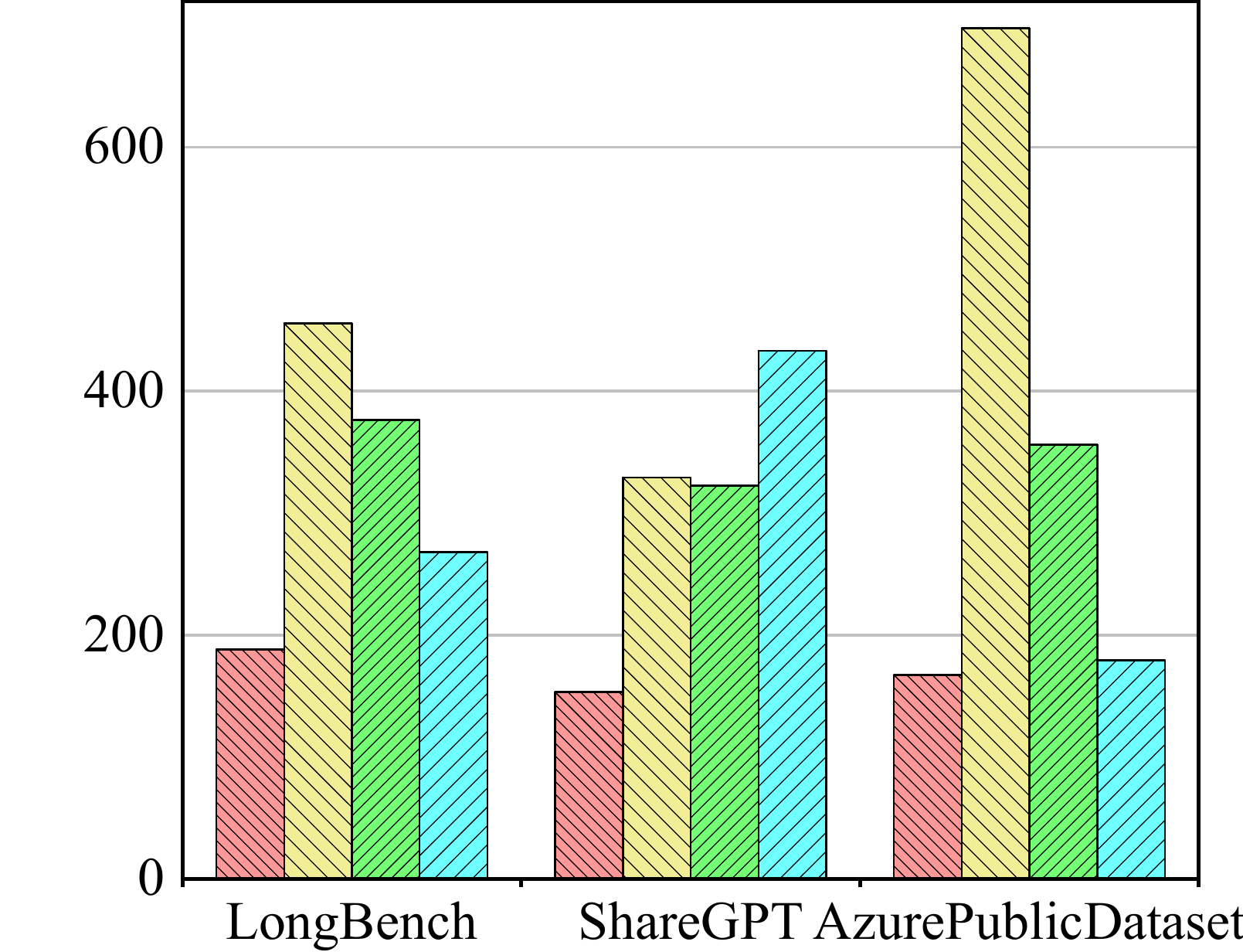}
        \caption{OPT-13B.}
    \end{subfigure}
    \caption{P99 TPOT on application workloads.} 
    \label{open_dataset_tpot}
\end{figure}

Besides the experiments driven by synthetic workloads, we evaluate the decoding latency by LongBench, ShareGPT and AzurePublicDataset as well. The experimental results presented in Figure~\ref{open_dataset_tpot} demonstrate that our framework remains the superiority over the others on application workloads. Specifically, AlignedServe reduces P99 TPOT latency by a maximum of $2.1\times$, $1.65 \times$, $7.4 \times$ on LongBench, ShareGPT and AzurePublicDataset, respectively. The reduced per-iteration latency provides a straightforward explanation for the higher throughput observed in Subsection~\ref{section4.2}.

\subsection{Ablation Study} \label{subsection4.4}

Subsection~\ref{section4.3} demonstrates that our framework achieves much lower latency compared with the others. In this subsection, we further give an ablation study to the latency of each iteration. Generally, an iteration in the decoding can be divided into two stages, i.e., iteration preparation and forward computing. In the period of iteration preparation, the scheduler may evict a request out of the batch, or add a new request to the batch. Both the two cases require the migration of KVCache, thus introduce additional latency. In the period of forward computing, the additional latency is mostly introduced by iteration-level bubbles. This subsection gives a deep insight into the two stages, respectively.

The latency involved in iteration preparation can be attributed to two aspects, i.e., the time to generate a new batch and the time to schedule an iteration. We analyze both the two type of latencies one by one as follows.

\textbf{The time to generate a new batch.} As discussed in Subsection~\ref{subsection3.4}, we propose the batch-level scheduling policy to issue requests to the decoding instances. Before a batch is scheduled, it must have already been generated by the prefix-aware batching policy, where generating the batch inevitably introduces additional latency. In our implementation, we pipeline the batch generation and decoding. When one batch is being decoded, the next candidate batch has already been generated and prefetched into the Candidate Batch Buffer. In this way, the additional latency introduced by batch generation is hidden. 
\begin{figure}[t]
    \begin{subfigure}[t]{0.3\linewidth}
        \centering
        \includegraphics[width=\textwidth]{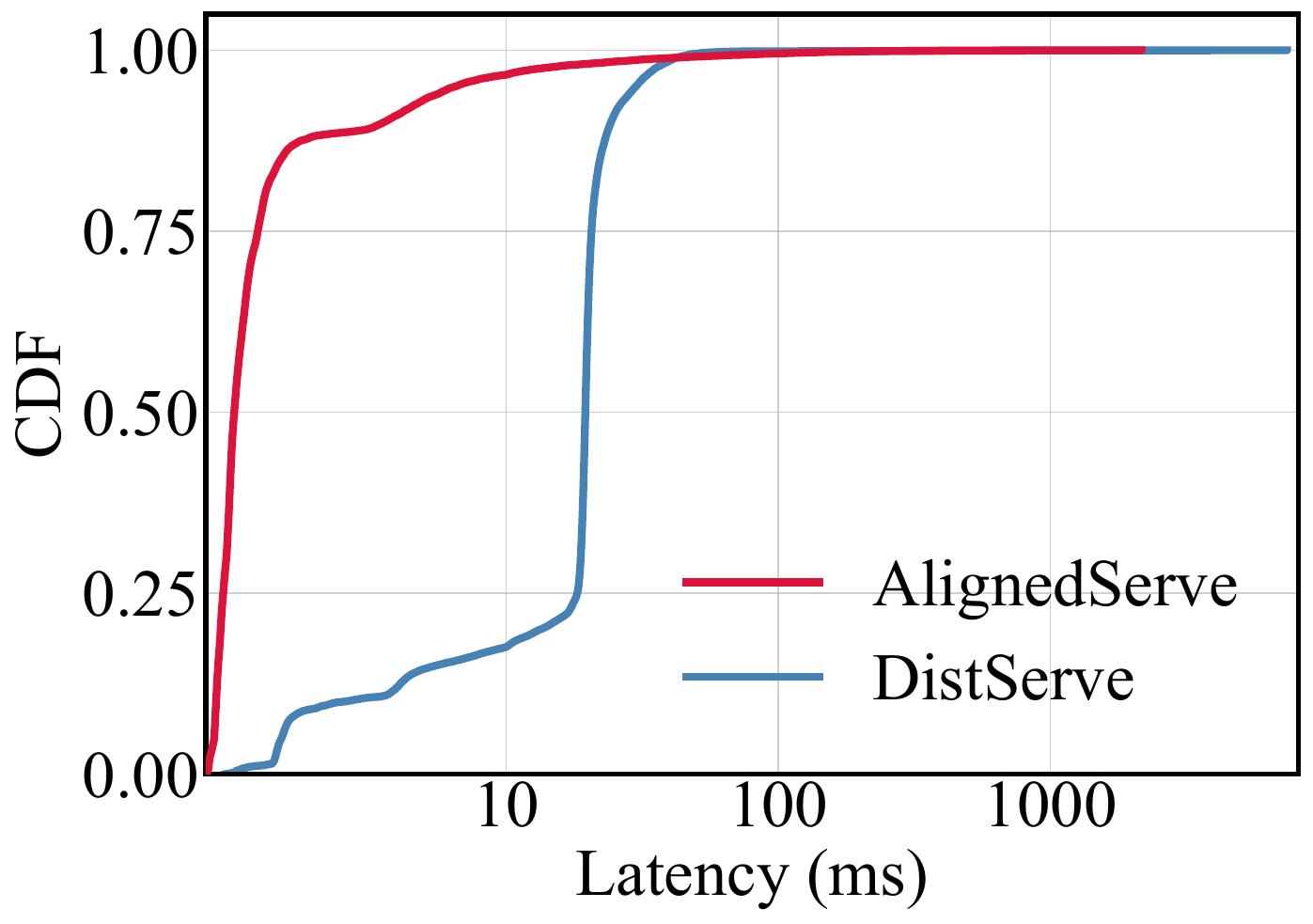} 
        \caption{ShareGPT.}
    \end{subfigure}%
    \begin{subfigure}[t]{0.3\linewidth}
        \centering
        \includegraphics[width=\textwidth]{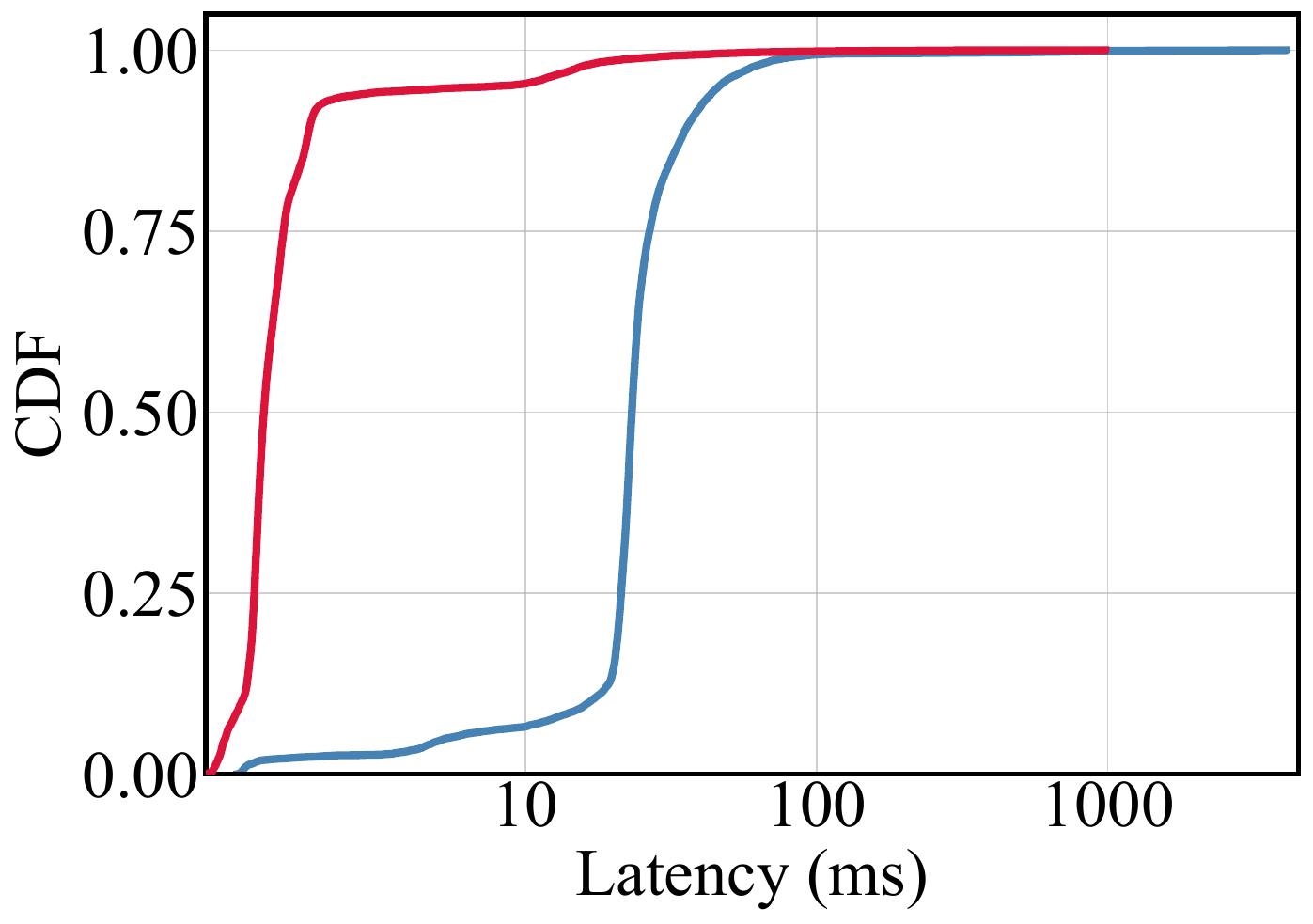} %
        \caption{LongBench.}
    \end{subfigure}
    \caption{The overhead of iteration scheduling.} 
    \label{cdf_for_step_latency}
\end{figure}

\begin{figure*}[t] 
    \centering 
    \begin{subfigure}{\textwidth}
        \centering
        \includegraphics[width=0.8\textwidth]{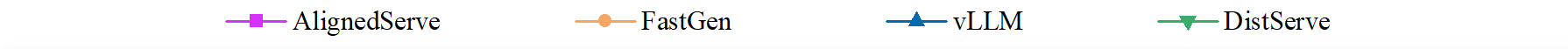}
    \end{subfigure}
    \begin{subfigure}[t]{0.248\textwidth} 
        \centering
        \includegraphics[width=\textwidth]{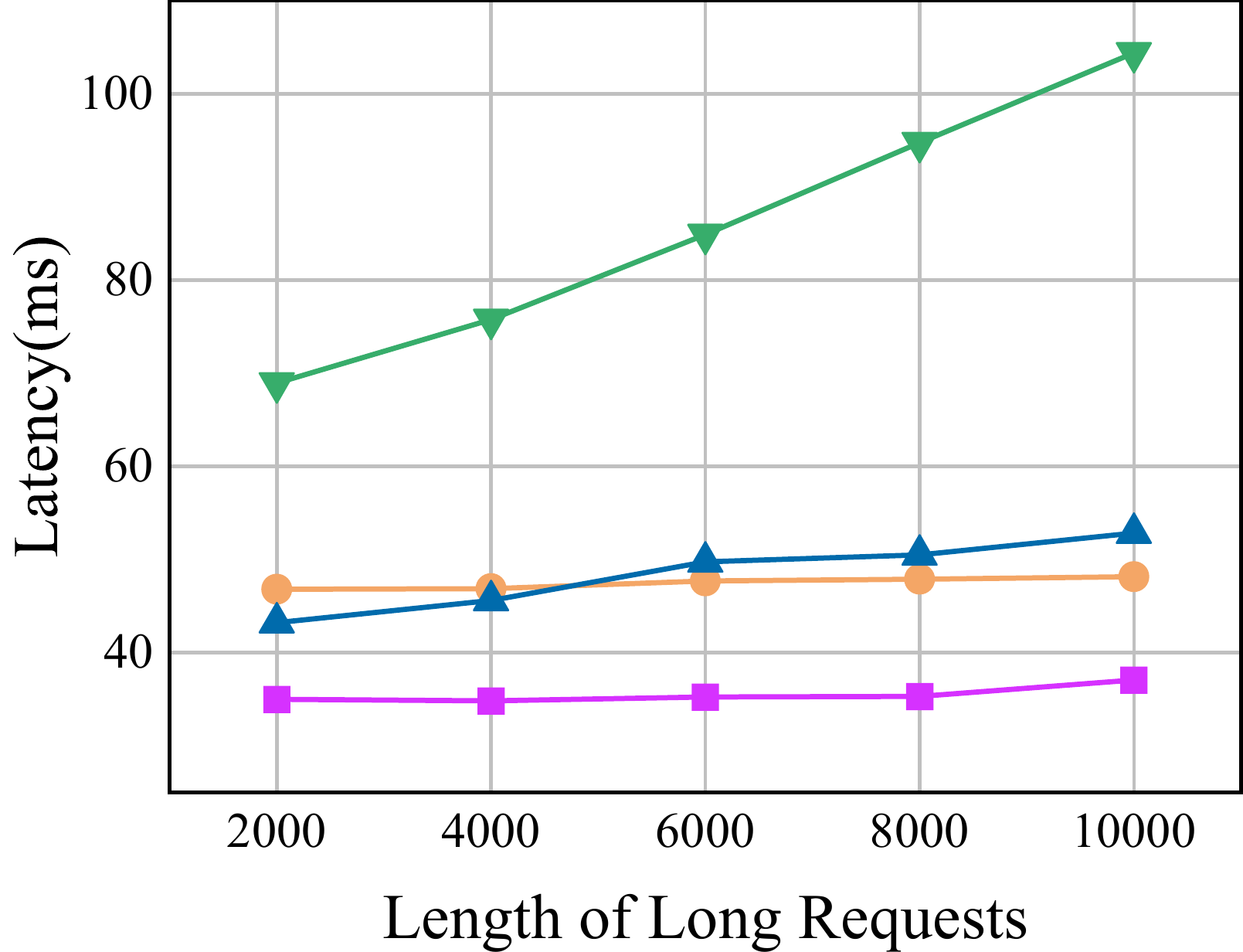}
        \caption{OPT-2.7B.}
    \end{subfigure}%
    \begin{subfigure}[t]{0.248\textwidth}
        \centering
        \includegraphics[width=\textwidth]{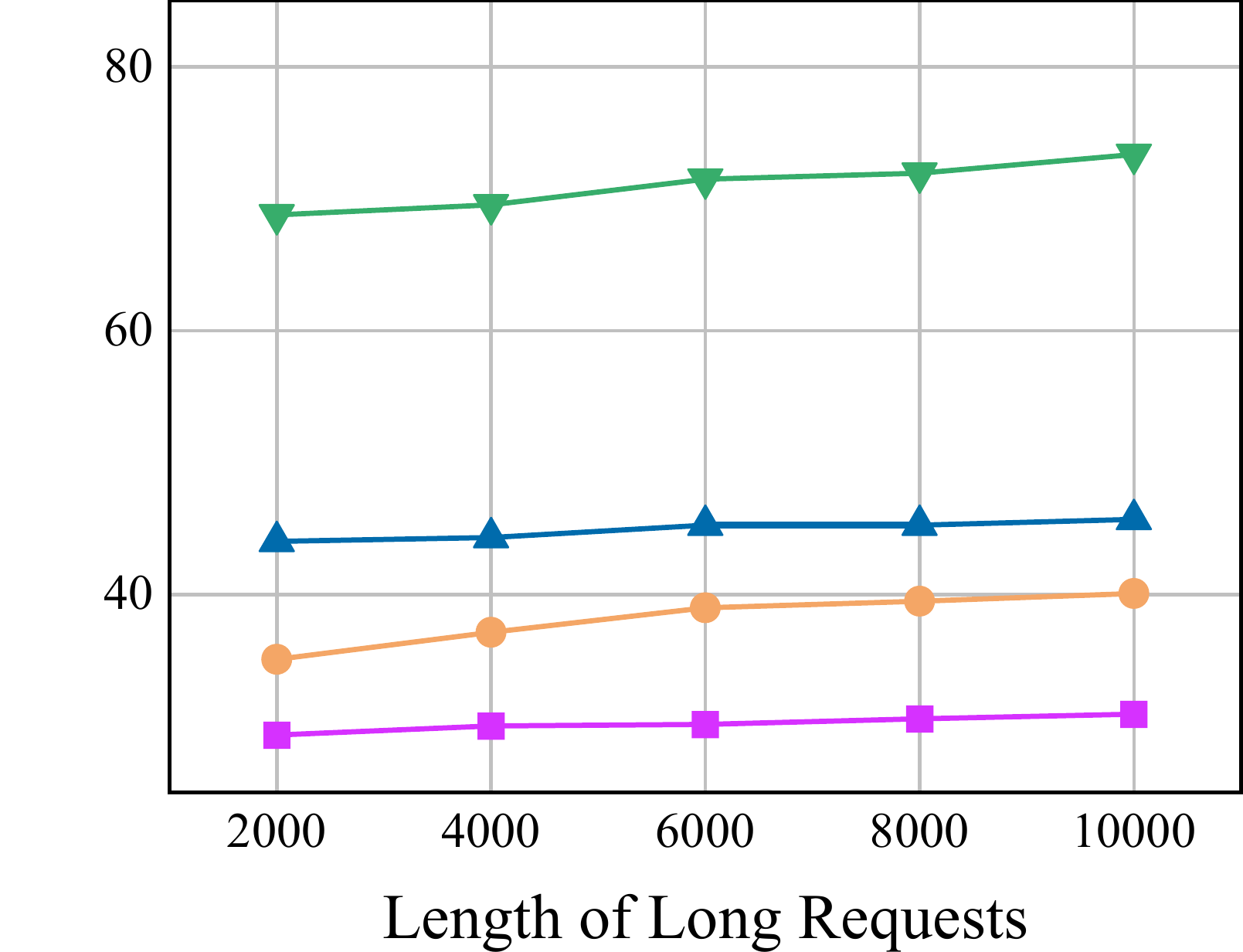}
        \caption{OPT-6.7B.}
    \end{subfigure}
    \begin{subfigure}[t]{0.248\textwidth}
        \centering
        \includegraphics[width=\textwidth]{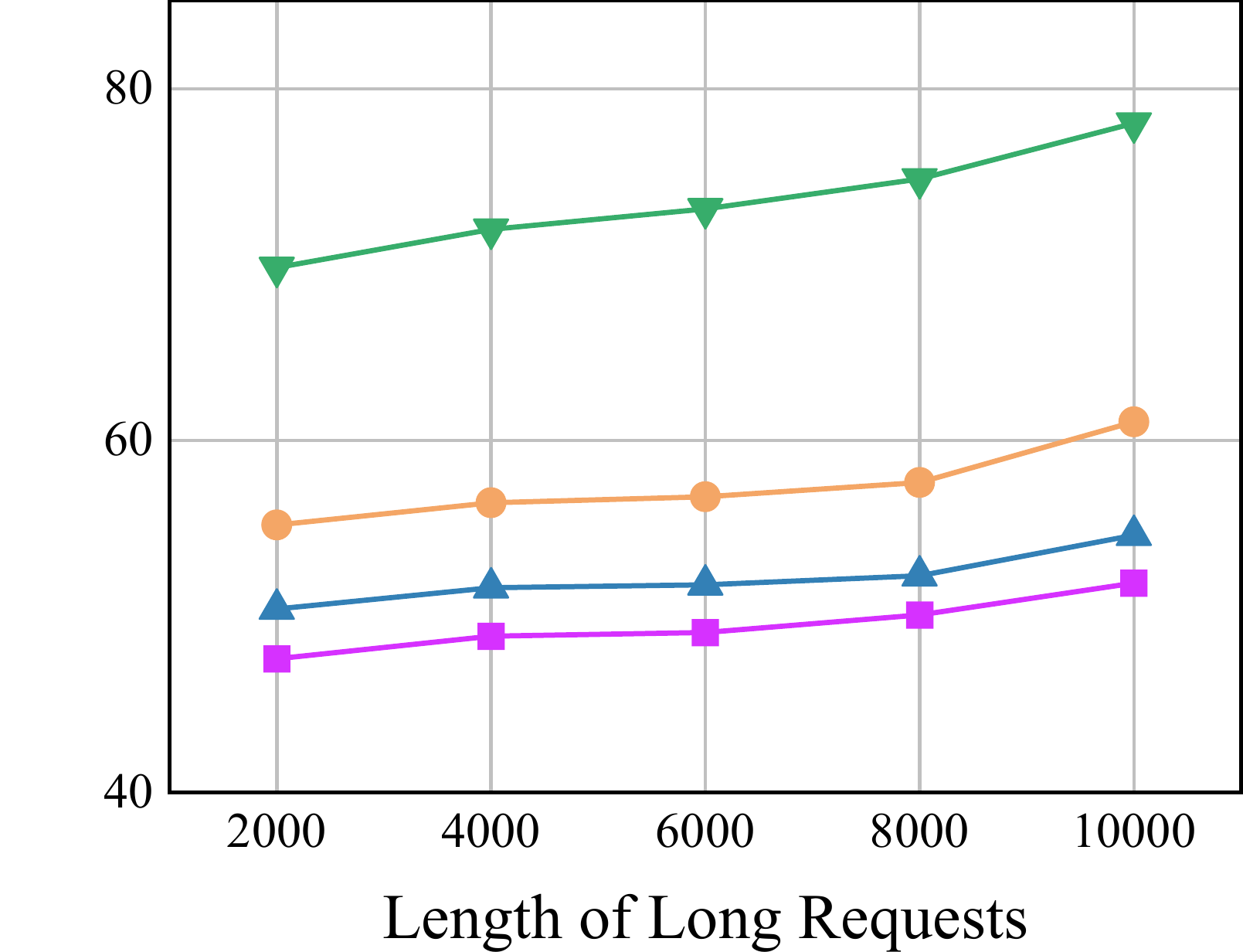}
        \caption{OPT-13B.}
    \end{subfigure}%
    \begin{subfigure}[t]{0.248\textwidth}
        \centering
        \includegraphics[width=\textwidth]{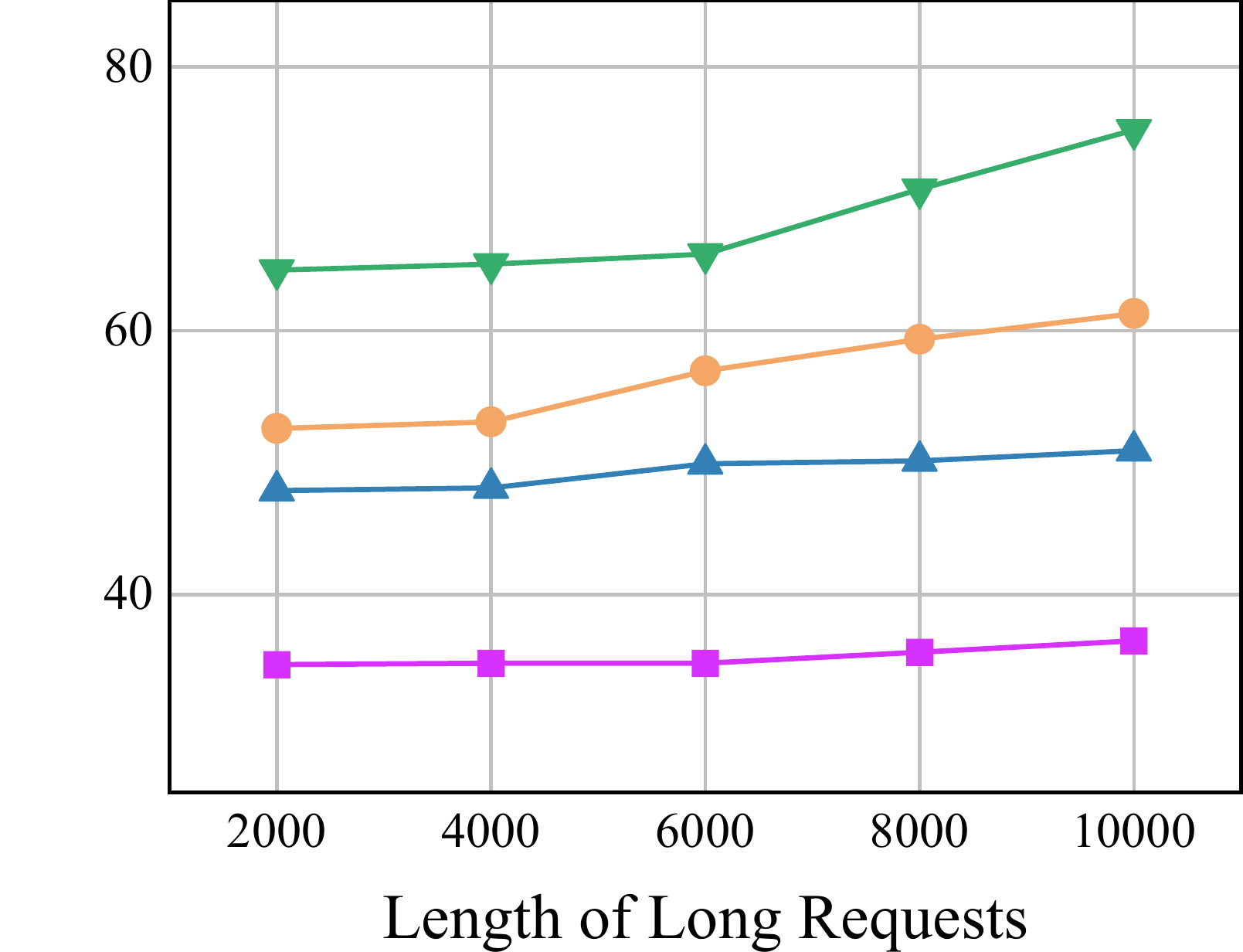}
        \caption{OPT-30B.}
    \end{subfigure}
    \caption{Comparison about the latency of forward computing in each iteration on synthetic workloads.}
    \label{Com-latency}
    
\end{figure*}

\textbf{The time to schedule an iteration.} When an iteration finishes, the scheduler prepares for the next iteration immediately. In this period of time, the scheduler may evict a request out of the batch due to that there is not enough memory to accommodate the tokens generated in the next iteration; or add a new request to the batch due to that some requests has finished. Both the two cases require to migrate KVCache, thus introduce additional latency. To evaluate this type of overhead, we conduct experiments to compare AlignedServe with DistServe in terms of time to schedule an iteration. The experiments run the ShareGPT and LongBench workloads on the OPT-6.7B model. Experimental results are shown in Figure~\ref{cdf_for_step_latency}. As the CDF of time to schedule an iteration shown in Figure~\ref{cdf_for_step_latency} presents,  more than 95\% of iterations in AlignedServe are scheduled within 5 ms, in contrast to DistServe, where 80\% of iterations take more than 10 ms.

\textbf{The latency of forward computing}. The latency involved in forward computing can be attributed to two aspects, i.e., the time to compute MLP and the time to compute attention, where the time to compute MLP is constant for a given model, but the time to compute attention varies significantly depending on the relied prefix of each token. As discussed in Subsection~\ref{section2.3}, mixing some long requests with these short ones leads to iteration-level bubbles as well as additional latency, which is the motivation of our prefix-aware batching policy. To evaluate the effectiveness of our proposal, we conduct experiments to compare the latency of forward computing in each iteration among different serving systems. 

In these experiments, we generate synthetic workloads whose long requests account for 95\%. Within such mixed workloads, the length of short requests remains constant, while the length of long requests increases gradually to simulate varied computing demands. Specifically, the lengths of long requests within the five workloads are 2000, 4000, 6000, 8000, and 10000, respectively. We replay these workloads on four different OPT models, and measure the latency of forward computing in each iteration. The experimental results are presented in Figure~\ref{Com-latency}. As shown in the figure, AlignedServe consistently achieves the lowest latency compared with the baselines. Furthermore, as the length of long requests increases from 2000 to 10000, the latency achieved by the counterparts (especially DistServe) increases remarkably, whereas the latency achieved by AlignedServe increases slightly. This observation indicates that AlignedServe is able to handle workloads composed of mixed short and long requests.

\begin{figure}[t]
    \begin{subfigure}[t]{0.3\linewidth}
        \centering
        \includegraphics[width=\textwidth]{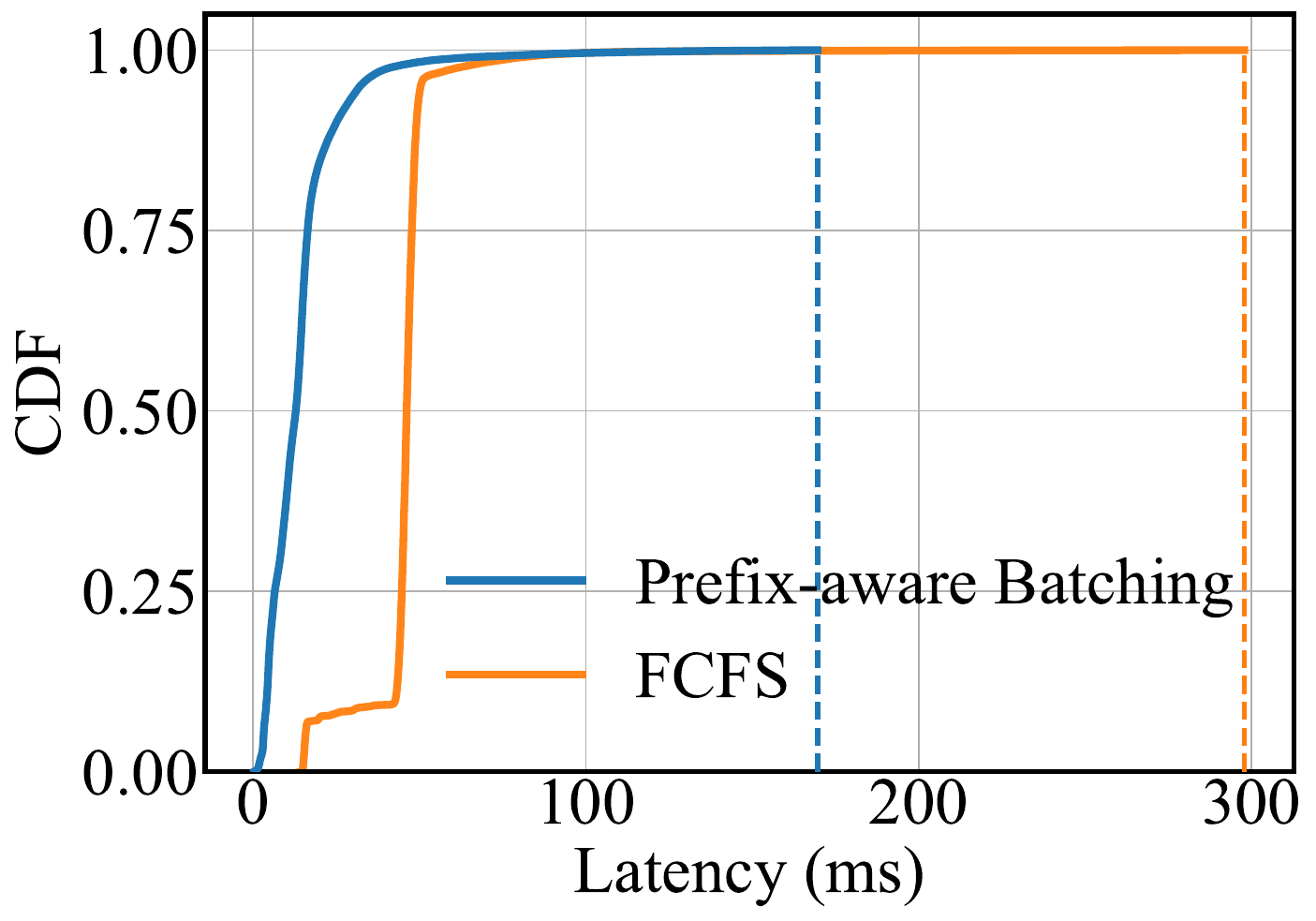}
        \caption{ShareGPT.}
    \end{subfigure}%
    \begin{subfigure}[t]{0.3\linewidth}
        \centering
        \includegraphics[width=\textwidth]{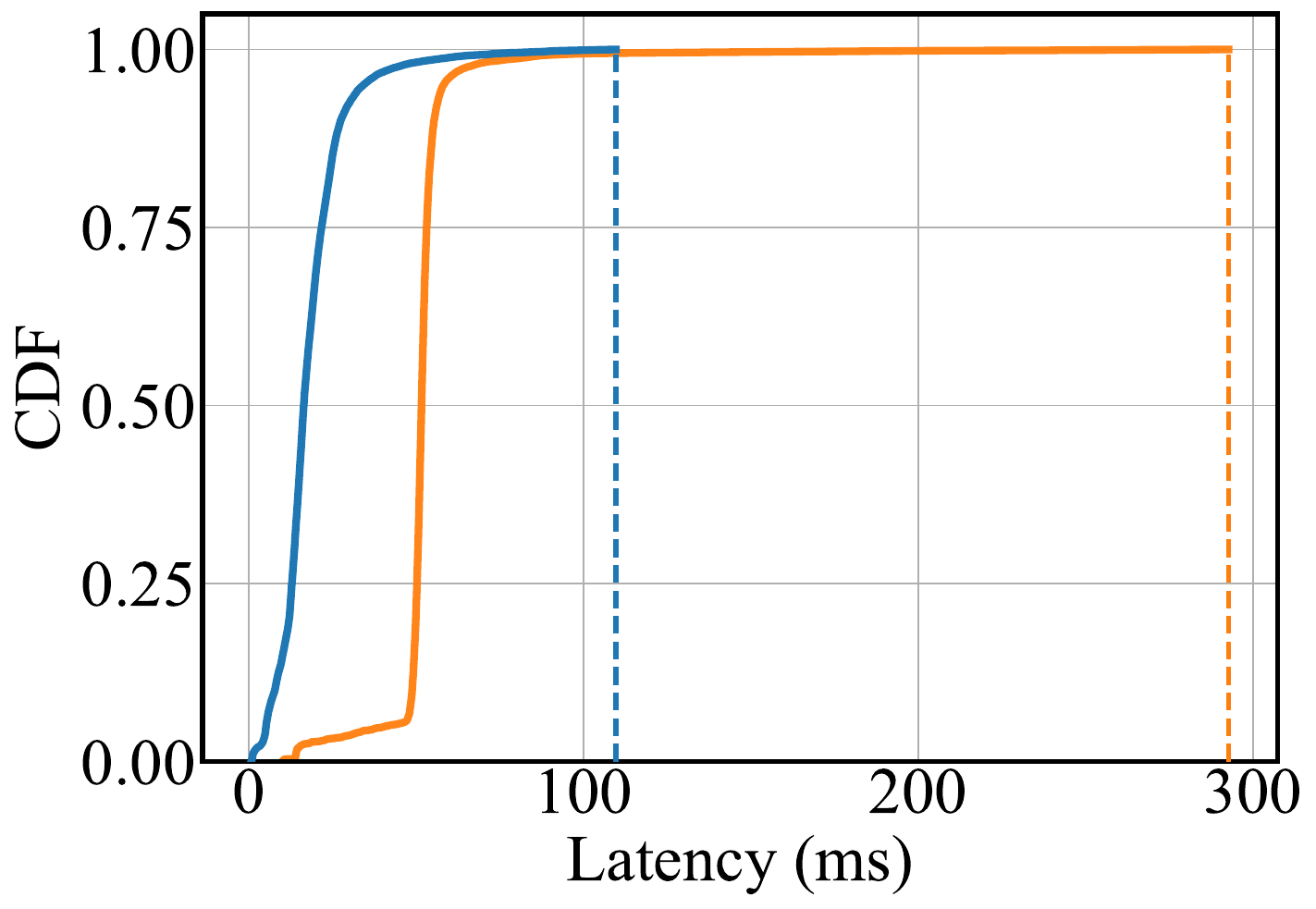}
        \caption{LongBench.}
    \end{subfigure}

    \caption{Comparison between our prefix-aware batching policy and FCFS in terms of latency involved in forward computing. AlignedServe achieves lower latency, indicating that our prefix-aware batching policy is effective.} 
    \label{cdf_step_latency}
\end{figure}

In order to demonstrate the effectiveness of our prefix-aware batching policy, we conduct experiments to compare our policy with the FCFS manner. Specifically, we replace our prefix-aware batching policy with FCFS in our framework, and compare the two counterparts using the OPT-13B model on realistic application workloads, i.e., LongBench and ShareGPT. The performance is also measured by the latency of forward computing. Experimental results are presented in Figure~\ref{cdf_step_latency}, which compares the CDF of the latency involved in forward computing between our prefix-aware batching policy and FCFS. From the figure we conclude that, under the prefix-aware batching policy, more than 90\% of iterations are able to complete the forward computing within 30ms, whereas the proportion of iterations that complete the forward computing within 30ms is less than 10\% for the FCFS policy.

\begin{figure}[t]
    \begin{subfigure}[t]{0.3\linewidth}
        \centering
        \includegraphics[width=\textwidth]{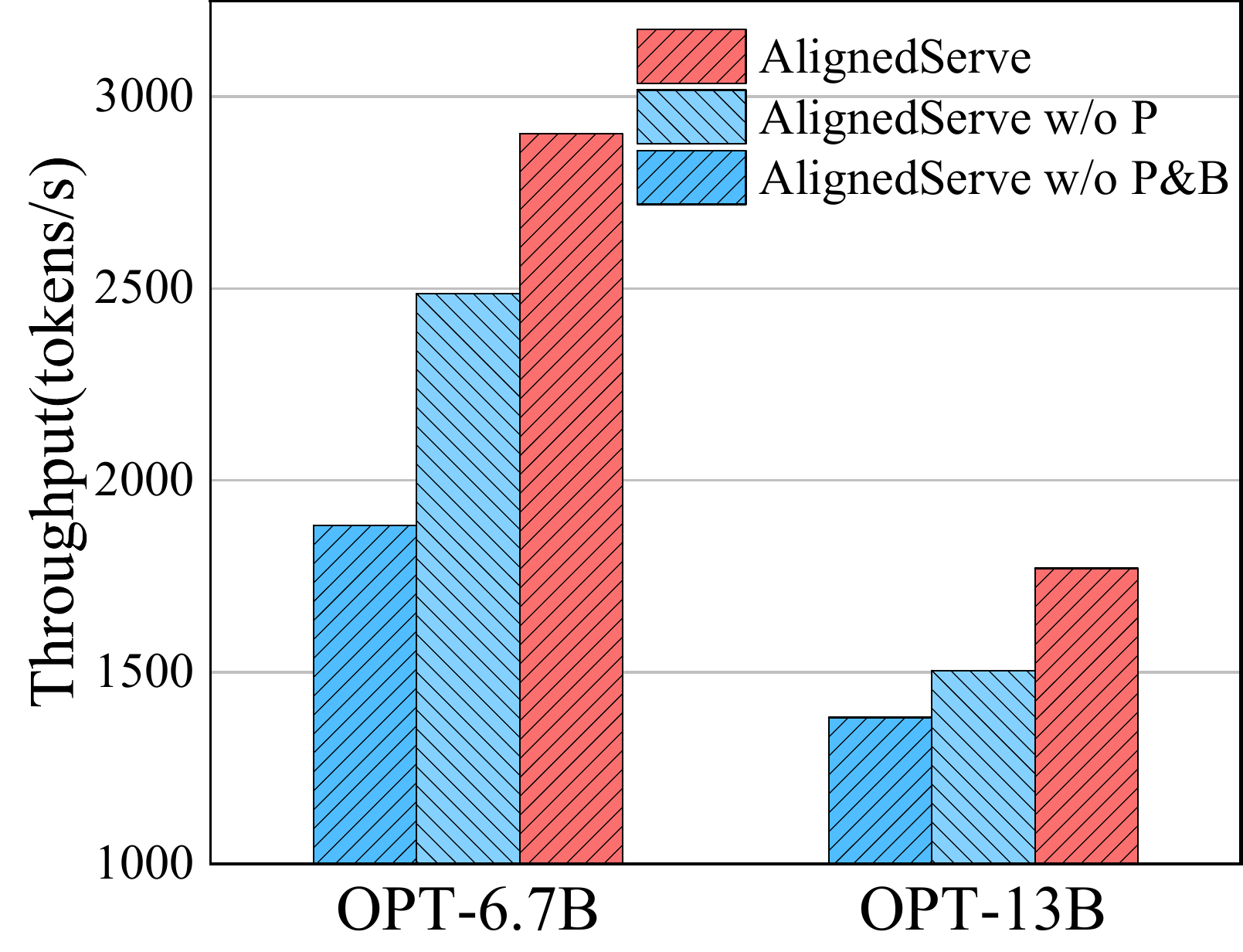}
        \caption{AzurePublicDataset.}
        \label{fig:AzurePublicDataset.}
    \end{subfigure}%
    \begin{subfigure}[t]{0.3\linewidth}
        \centering
        \includegraphics[width=\textwidth]{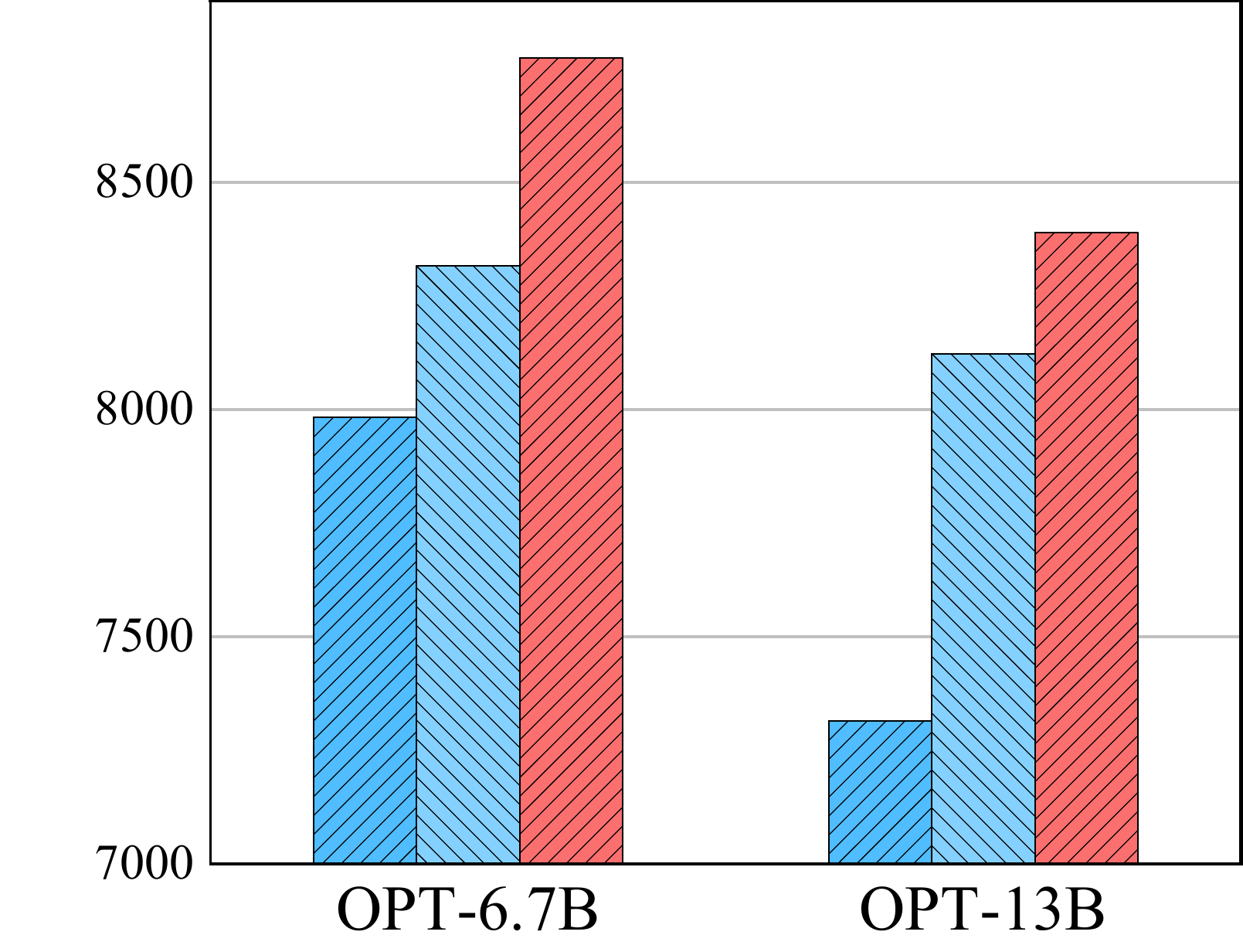}
        \caption{LongBench.}
    \end{subfigure}
    \caption{Ablation of prefetching and prefix-aware batching.} 
    \label{abl_prefetch}
\end{figure}

\textbf{Ablation study about GPU prefetching and prefix-aware batching.} We evaluate the three comparison candidates, i.e., AlignedServe, AlignedServe without GPU prefetching (AlignedServe w/o P), AlignedServe without both GPU prefetching and prefix-aware batching (AlignedServe w/o P\&B). The experimental results shown in Figure~\ref{abl_prefetch} demonstrate that, the throughput decreases by 14.73\% when GPU Prefetching is disabled, and decreases by 28.51\% further when both the two optimizations are disabled on AzureDataset.

\begin{figure}[t]
    \begin{subfigure}[t]{0.3\linewidth}
        \centering
        \includegraphics[width=\textwidth]{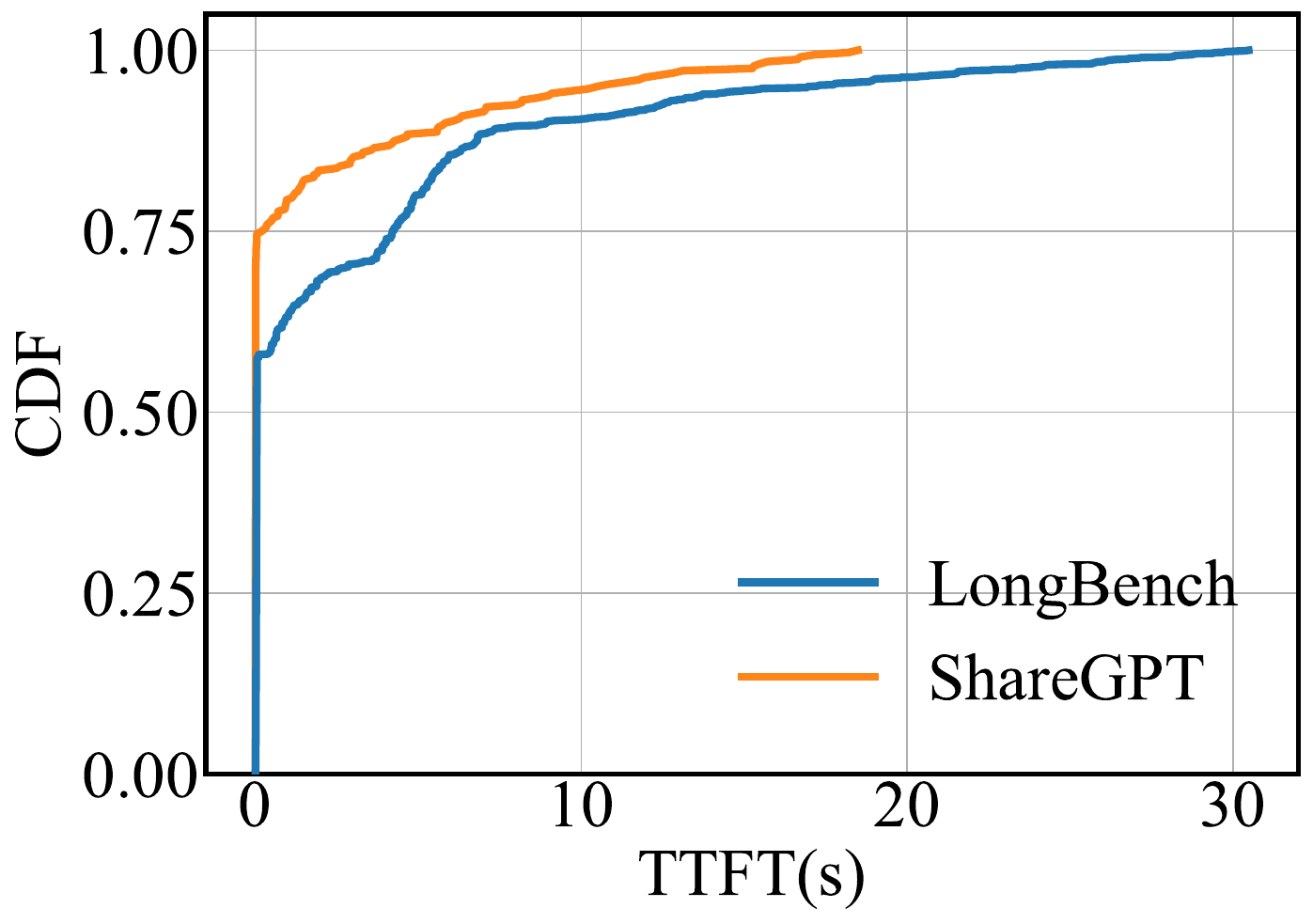}
        \caption{OPT-6.7B.}
    \end{subfigure}%
    \begin{subfigure}[t]{0.3\linewidth}
        \centering
        \includegraphics[width=\textwidth]{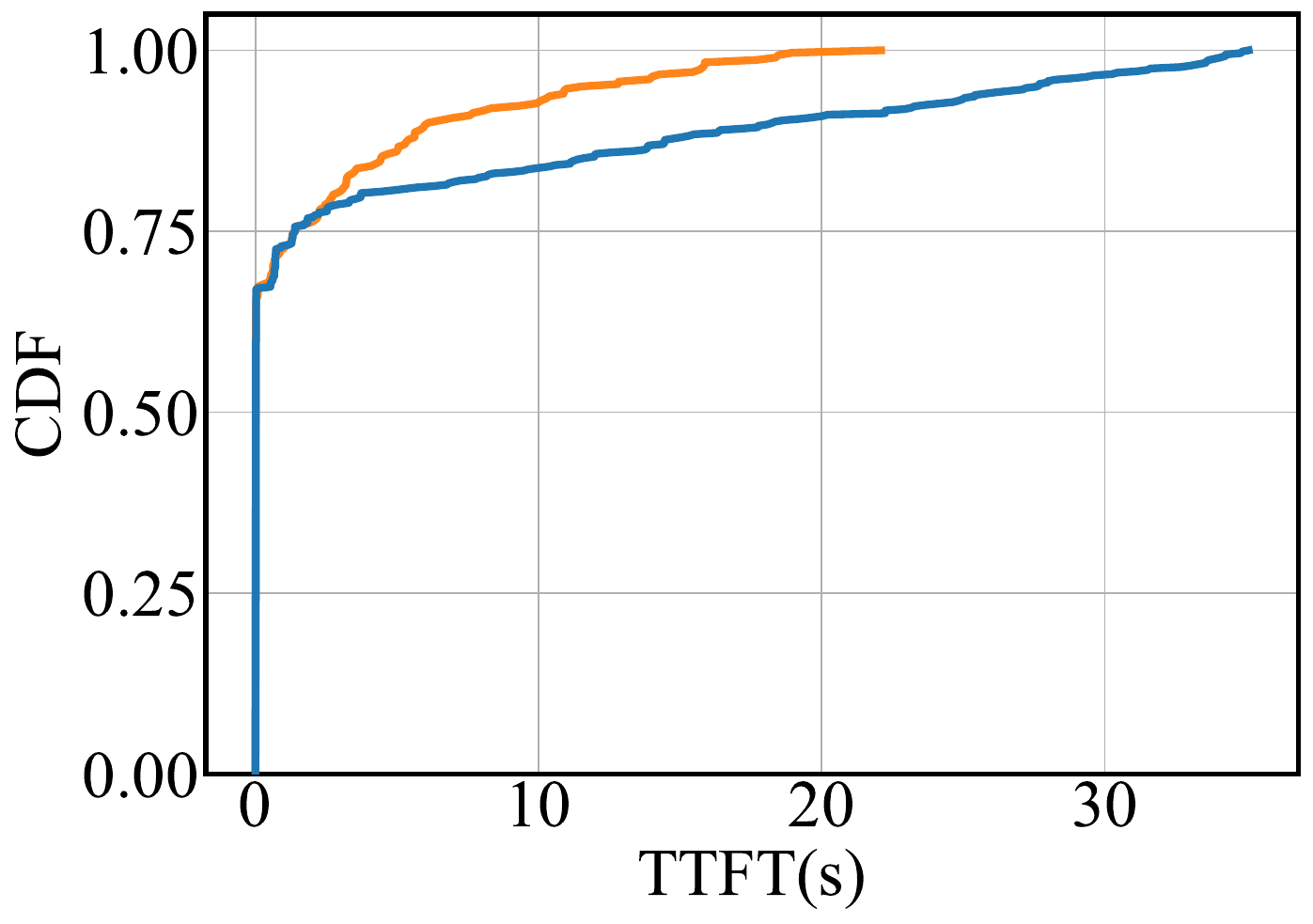}
        \caption{OPT-13B.}
    \end{subfigure}

    \caption{CDF of TTFT.} 
    \label{open_dataset_ttft}
\end{figure}

\textbf{Overhead of batch switch.} When the batch switch occurs, requests belonging to different batches coexist in the running batch, indicating that our primary design philosophy (i.e., serving requests with the similar length of prefix together) is violated. However, we argue that the period of time used for batch switch is limited. Specifically, we conducted experiments to evaluate how many iterations that the batch switch is occurring under the OPT-6.7B model. Experimental results demonstrate that the fraction of iterations that contain requests from different batches is no more than 8.61\% and 12.37\% on ShareGPT and LongBench, respectively.

\textbf{Overhead of KV pool.} AlignedServe offloads large volume of KVCache into CPU memory. In our experiments, the KV pool is set to be $800$GB. However, the statistics derived from the experiments demonstrates that the actual consumption ranges from $20$GB to $250$GB under the OPT-6.7B model on LongBench. 

\textbf{TTFT.} Our Prefix-aware Batching policy introduces additional latency thus deteriorates the TTFT. To evaluate this negative impact, we disable the starvation handling mechanism and measure the TTFT across different workloads. As Figure~\ref{open_dataset_ttft} shows, the average TTFTs for ShareGPT and LongBench are 1.49s and 2.54s, respectively, which are acceptable in production. Even though the maximum TTFT is about 30s, if the starvation handling mechanism is retrieved, the TTFT can be adjusted on demand.

\section{Related Works}

\textbf{Disaggregated Architecture.} Disaggregated architecture has been adopted by many inference frameworks~\cite{zhong2024distserve,patel2024splitwise,305212,hu2024inference}. Specifically, the disaggregated inference architecture decouples the prefill (compute-bound) and decode (memory-bound) phases, allowing hardware allocation and parallelization strategies to be optimized independently, so as to improve TTFT and TPOT, without phase interference. LoongServe~\cite{wu2024loongserve} further proposes to handle dynamic workloads by integrating elastic scaling mechanisms. MemServe~\cite{hu2024memserve} combines context caching with disaggregated inference.\par
\textbf{LLM Inference batching and scheduling Optimizations.}
Many studies focus on the batching and scheduling~\cite{holmes2024deepspeed, wu2023fast} of LLM inference. Orca~\cite{orca} is the first work explicitly considering the varied input and output lengths, and accordingly changes the inference scheduling from traditional request-level to the fine-grained iteration-level. Sarathi-Serve~\cite{sarathi-serve} divides input prompts into fixed-size chunks and mixes these chunks with decodes to reduce the pipeline bubbles caused by the warm-up stages of varied-length input requests. Apt-serve~\cite{apt-serve} combines KVCache with a memory-efficient hidden cache to maintain much more reusable input hidden state vectors, and further employs an adaptive runtime scheduling mechanism that dynamically optimizes batch composition, thus improves the throughput significantly. These state-of-the-art systems are all orthogonal to AlignedServe. \par

\textbf{KVCache Management.} Employing the KVCache to reduce redundant computing is an efficient mechanism used to optimize TTFT. PagedAttention~\cite{vllm} draws inspiration from traditional operating system concepts such as paging and virtual memory, allowing KVCache to be non-contiguously allocated and logically organized during execution, so as to alleviate memory capacity pressure. InfiniGen~\cite{298683}, FlexGen~\cite{10.5555/3618408.3619696} and CachedAttention~\cite{10.5555/3691992.3691999} explore tiered KVCache by offloading cold \textit{KV}s to host DRAM or even SSDs. 
As long-context inference has attracted much attention, a large number of works focus on KVCache management of the shared prefix. The state-of-the-art works such as HotPrefix~\cite{10.1145/3749168}, BatchLLM~\cite{zheng2025batchllmoptimizinglargebatched}, Cache-Craft~\cite{10.1145/3725273}, and Preble~\cite{srivatsa2025preble} fall into this category. The varied lengths of prefixes generated by these works can be fed into AlignedServe further to achieve high throughput. And our GPU-Prefetch-For-GPU architecture can be borrowed by these works to accelerate the transmission of KVCache between GPUs and CPUs.

\section{Conclusion}

Traditional LLM serving systems mostly neglected the fact that the inference requests are of different lengths, where grouping different lengths of requests into a batch leads to iteration-level bubbles. This work proposes a novel LLM serving framework called AlignedServe, which employs the prefix-aware batching policy that groups requests with similar length of prefix into a batch, thereby significantly eliminating iteration-level bubbles. To efficiently support the proposed prefix-aware batching policy, we further design a novel disaggregated architecture as well as the corresponding batch-level scheduling policy to reduce the scheduling overheads. The experiments driven by varied workloads demonstrate that our AlignedServe achieves higher performance (thus high-throughput) and lower scheduling overhead (thus computing-efficient) compared with the state-of-the-art works.

\begin{acks}
We would like to thank the anonymous reviewers for their valuable suggestions on improving the presentation of the paper. This work is supported by Guangdong S\&T Program under Grant No. 2025B0101080001, the National Natural Science Foundation of China (NSFC) under Grant No. 62272499 and No. 62332021, the Guangdong Province Special Support Program for Cultivating High-Level Talents under Grant No. 2021T006X160, and PazhouLab under Grant No. PZL2023KF0001.
\end{acks}

\bibliographystyle{ACM-Reference-Format}
\bibliography{sample-base}
\appendix

\end{document}